\documentclass[twocolumn]{aastex62}

\usepackage{textcomp}

\reportnum{DES-2018-0403}
\reportnum{FERMILAB-PUB-18-621-AE}

\shorttitle{Continuum Reverberation Mapping on DES}
\shortauthors{Yu et al.}
\begin{document}

\title{Quasar Accretion Disk Sizes from Continuum Reverberation Mapping in the DES Standard Star Fields}

\author[0000-0003-0644-9282]{Zhefu Yu}
\affiliation{Department of Astronomy, The Ohio State University, Columbus, Ohio 43210, USA}

\author[0000-0002-4279-4182]{Paul Martini}
\affiliation{Department of Astronomy, The Ohio State University, Columbus, Ohio 43210, USA}
\affiliation{Center of Cosmology and Astro-Particle Physics, The Ohio State University, Columbus, Ohio, 43210, USA}

\author[0000-0002-4213-8783]{T.~M.~Davis}
\affiliation{School of Mathematics and Physics, University of Queensland,  Brisbane, QLD 4072, Australia}
\affiliation{ARC Centre of Excellence for All-sky Astrophysics (CAASTRO), Australia}

\author{R.~A.~Gruendl}
\affiliation{Department of Astronomy, University of Illinois at Urbana-Champaign, 1002 W. Green Street, Urbana, IL 61801, USA}
\affiliation{National Center for Supercomputing Applications, 1205 West Clark St., Urbana, IL 61801, USA}

\author{J.~K.~Hoormann}
\affiliation{School of Mathematics and Physics, University of Queensland,  Brisbane, QLD 4072, Australia}

\author[0000-0001-6017-2961]{C.~S.~Kochanek}
\affiliation{Department of Astronomy, The Ohio State University, Columbus, Ohio 43210, USA}
\affiliation{Center of Cosmology and Astro-Particle Physics, The Ohio State University, Columbus, Ohio, 43210, USA}

\author[0000-0003-1731-0497]{C.~Lidman}
\affiliation{ARC Centre of Excellence for All-sky Astrophysics (CAASTRO), Australia}
\affiliation{Australian Astronomical Observatory, North Ryde, NSW 2113, Australia}

\author[0000-0003-2371-4121]{D.~Mudd}
\affiliation{Department of Physics and Astronomy, University of California, Irvine, Irvine, CA 92697, USA}

\author{B.~M.~Peterson}
\affiliation{Department of Astronomy, The Ohio State University, Columbus, Ohio 43210, USA}
\affiliation{Center of Cosmology and Astro-Particle Physics, The Ohio State University, Columbus, Ohio, 43210, USA}
\affiliation{Space Telescope Science Institute, 3700 San Martin Drive, Baltimore, MD 21218, USA}

\author{W.~Wester}
\affiliation{Fermi National Accelerator Laboratory, P. O. Box 500, Batavia, IL 60510, USA}

\author[0000-0002-7069-7857]{S.~Allam}
\affiliation{Fermi National Accelerator Laboratory, P. O. Box 500, Batavia, IL 60510, USA}

\author[0000-0002-0609-3987]{J.~Annis}
\affiliation{Fermi National Accelerator Laboratory, P. O. Box 500, Batavia, IL 60510, USA}

\author{J.~Asorey}
\affiliation{Korea Astronomy and Space Science Institute, Yuseong-gu, Daejeon, 305-348, Korea}

\author{S.~Avila}
\affiliation{Institute of Cosmology and Gravitation, University of Portsmouth, Portsmouth, PO1 3FX, UK}

\author{M.~Banerji}
\affiliation{Kavli Institute for Cosmology, University of Cambridge, Madingley Road, Cambridge CB3 0HA, UK}
\affiliation{Institute of Astronomy, University of Cambridge, Madingley Road, Cambridge CB3 0HA, UK}

\author{E.~Bertin}
\affiliation{Sorbonne Universit\'es, UPMC Univ Paris 06, UMR 7095, Institut d'Astrophysique de Paris, F-75014, Paris, France}
\affiliation{CNRS, UMR 7095, Institut d'Astrophysique de Paris, F-75014, Paris, France}

\author{D.~Brooks}
\affiliation{Department of Physics \& Astronomy, University College London, Gower Street, London, WC1E 6BT, UK}

\author[0000-0002-3304-0733]{E.~Buckley-Geer}
\affiliation{Fermi National Accelerator Laboratory, P. O. Box 500, Batavia, IL 60510, USA}

\author{J.~Calcino}
\affiliation{School of Mathematics and Physics, University of Queensland,  Brisbane, QLD 4072, Australia}

\author[0000-0003-3044-5150]{A.~Carnero~Rosell}
\affiliation{Laborat\'orio Interinstitucional de e-Astronomia - LIneA, Rua Gal. Jos\'e Cristino 77, Rio de Janeiro, RJ - 20921-400, Brazil}
\affiliation{Centro de Investigaciones Energ\'eticas, Medioambientales y Tecnol\'ogicas (CIEMAT), Madrid, Spain}

\author{D.~Carollo}
\affiliation{ARC Centre of Excellence for All-sky Astrophysics (CAASTRO), Australia}
\affiliation{INAF, Astrophysical Observatory of Turin, Torino, Italy}

\author[0000-0002-4802-3194]{M.~Carrasco~Kind}
\affiliation{National Center for Supercomputing Applications, 1205 West Clark St., Urbana, IL 61801, USA}
\affiliation{Department of Astronomy, University of Illinois at Urbana-Champaign, 1002 W. Green Street, Urbana, IL 61801, USA}

\author[0000-0002-3130-0204]{J.~Carretero}
\affiliation{Institut de F\'{\i}sica d'Altes Energies (IFAE), The Barcelona Institute of Science and Technology, Campus UAB, 08193 Bellaterra (Barcelona) Spain}

\author{C.~E.~Cunha}
\affiliation{Kavli Institute for Particle Astrophysics \& Cosmology, P. O. Box 2450, Stanford University, Stanford, CA 94305, USA}

\author[0000-0002-8198-0332]{C.~B.~D'Andrea}
\affiliation{Department of Physics and Astronomy, University of Pennsylvania, Philadelphia, PA 19104, USA}

\author{L.~N.~da Costa}
\affiliation{Laborat\'orio Interinstitucional de e-Astronomia - LIneA, Rua Gal. Jos\'e Cristino 77, Rio de Janeiro, RJ - 20921-400, Brazil}
\affiliation{Observat\'orio Nacional, Rua Gal. Jos\'e Cristino 77, Rio de Janeiro, RJ - 20921-400, Brazil}

\author[0000-0001-8318-6813]{J.~De~Vicente}
\affiliation{Centro de Investigaciones Energ\'eticas, Medioambientales y Tecnol\'ogicas (CIEMAT), Madrid, Spain}

\author[0000-0002-0466-3288]{S.~Desai}
\affiliation{Department of Physics, IIT Hyderabad, Kandi, Telangana 502285, India}

\author[0000-0002-8357-7467]{H.~T.~Diehl}
\affiliation{Fermi National Accelerator Laboratory, P. O. Box 500, Batavia, IL 60510, USA}

\author{P.~Doel}
\affiliation{Department of Physics \& Astronomy, University College London, Gower Street, London, WC1E 6BT, UK}

\author[0000-0002-1894-3301]{T.~F.~Eifler}
\affiliation{Department of Astronomy/Steward Observatory, 933 North Cherry Avenue, Tucson, AZ 85721-0065, USA}
\affiliation{Jet Propulsion Laboratory, California Institute of Technology, 4800 Oak Grove Dr., Pasadena, CA 91109, USA}

\author{B.~Flaugher}
\affiliation{Fermi National Accelerator Laboratory, P. O. Box 500, Batavia, IL 60510, USA}

\author{P.~Fosalba}
\affiliation{Institut d'Estudis Espacials de Catalunya (IEEC), 08034 Barcelona, Spain}
\affiliation{Institute of Space Sciences (ICE, CSIC),  Campus UAB, Carrer de Can Magrans, s/n,  08193 Barcelona, Spain}

\author[0000-0003-4079-3263]{J.~Frieman}
\affiliation{Kavli Institute for Cosmological Physics, University of Chicago, Chicago, IL 60637, USA}
\affiliation{Fermi National Accelerator Laboratory, P. O. Box 500, Batavia, IL 60510, USA}

\author[0000-0002-9370-8360]{J.~Garc\'ia-Bellido}
\affiliation{Instituto de Fisica Teorica UAM/CSIC, Universidad Autonoma de Madrid, 28049 Madrid, Spain}

\author{E.~Gaztanaga}
\affiliation{Institute of Space Sciences (ICE, CSIC),  Campus UAB, Carrer de Can Magrans, s/n,  08193 Barcelona, Spain}
\affiliation{Institut d'Estudis Espacials de Catalunya (IEEC), 08034 Barcelona, Spain}

\author{K.~Glazebrook}
\affiliation{Centre for Astrophysics \& Supercomputing, Swinburne University of Technology, Victoria 3122, Australia}

\author{D.~Gruen}
\affiliation{SLAC National Accelerator Laboratory, Menlo Park, CA 94025, USA}
\affiliation{Kavli Institute for Particle Astrophysics \& Cosmology, P. O. Box 2450, Stanford University, Stanford, CA 94305, USA}

\author{J.~Gschwend}
\affiliation{Observat\'orio Nacional, Rua Gal. Jos\'e Cristino 77, Rio de Janeiro, RJ - 20921-400, Brazil}
\affiliation{Laborat\'orio Interinstitucional de e-Astronomia - LIneA, Rua Gal. Jos\'e Cristino 77, Rio de Janeiro, RJ - 20921-400, Brazil}

\author[0000-0003-0825-0517]{G.~Gutierrez}
\affiliation{Fermi National Accelerator Laboratory, P. O. Box 500, Batavia, IL 60510, USA}

\author{W.~G.~Hartley}
\affiliation{Department of Physics \& Astronomy, University College London, Gower Street, London, WC1E 6BT, UK}
\affiliation{Department of Physics, ETH Zurich, Wolfgang-Pauli-Strasse 16, CH-8093 Zurich, Switzerland}

\author{S.~R.~Hinton}
\affiliation{School of Mathematics and Physics, University of Queensland,  Brisbane, QLD 4072, Australia}

\author{D.~L.~Hollowood}
\affiliation{Santa Cruz Institute for Particle Physics, Santa Cruz, CA 95064, USA}

\author{K.~Honscheid}
\affiliation{Center of Cosmology and Astro-Particle Physics, The Ohio State University, Columbus, Ohio, 43210, USA}
\affiliation{Department of Physics, The Ohio State University, Columbus, OH 43210, USA}

\author[0000-0002-2571-1357]{B.~Hoyle}
\affiliation{Max Planck Institute for Extraterrestrial Physics, Giessenbachstrasse, 85748 Garching, Germany}
\affiliation{Universit\"ats-Sternwarte, Fakult\"at f\"ur Physik, Ludwig-Maximilians Universit\"at M\"unchen, Scheinerstr. 1, 81679 M\"unchen, Germany}

\author{D.~J.~James}
\affiliation{Harvard-Smithsonian Center for Astrophysics, Cambridge, MA 02138, USA}

\author{A.~G.~Kim}
\affiliation{Lawrence Berkeley National Laboratory, 1 Cyclotron Road, Berkeley, CA 94720, USA}

\author{E.~Krause}
\affiliation{Department of Astronomy/Steward Observatory, 933 North Cherry Avenue, Tucson, AZ 85721-0065, USA}

\author[0000-0003-0120-0808]{K.~Kuehn}
\affiliation{Australian Astronomical Observatory, North Ryde, NSW 2113, Australia}

\author[0000-0003-2511-0946]{N.~Kuropatkin}
\affiliation{Fermi National Accelerator Laboratory, P. O. Box 500, Batavia, IL 60510, USA}

\author{G.~F.~Lewis}
\affiliation{Sydney Institute for Astronomy, School of Physics, A28, The University of Sydney, NSW 2006, Australia}

\author{M.~Lima}
\affiliation{Departamento de F\'isica Matem\'atica, Instituto de F\'isica, Universidade de S\~ao Paulo, CP 66318, S\~ao Paulo, SP, 05314-970, Brazil}
\affiliation{Laborat\'orio Interinstitucional de e-Astronomia - LIneA, Rua Gal. Jos\'e Cristino 77, Rio de Janeiro, RJ - 20921-400, Brazil}

\author{E.~Macaulay}
\affiliation{Institute of Cosmology and Gravitation, University of Portsmouth, Portsmouth, PO1 3FX, UK}

\author{M.~A.~G.~Maia}
\affiliation{Observat\'orio Nacional, Rua Gal. Jos\'e Cristino 77, Rio de Janeiro, RJ - 20921-400, Brazil}
\affiliation{Laborat\'orio Interinstitucional de e-Astronomia - LIneA, Rua Gal. Jos\'e Cristino 77, Rio de Janeiro, RJ - 20921-400, Brazil}

\author[0000-0003-0710-9474]{J.~L.~Marshall}
\affiliation{George P. and Cynthia Woods Mitchell Institute for Fundamental Physics and Astronomy, and Department of Physics and Astronomy, Texas A\&M University, College Station, TX 77843,  USA}

\author[0000-0002-1372-2534]{F.~Menanteau}
\affiliation{National Center for Supercomputing Applications, 1205 West Clark St., Urbana, IL 61801, USA}
\affiliation{Department of Astronomy, University of Illinois at Urbana-Champaign, 1002 W. Green Street, Urbana, IL 61801, USA}

\author[0000-0002-6610-4836]{R.~Miquel}
\affiliation{Institut de F\'{\i}sica d'Altes Energies (IFAE), The Barcelona Institute of Science and Technology, Campus UAB, 08193 Bellaterra (Barcelona) Spain}
\affiliation{Instituci\'o Catalana de Recerca i Estudis Avan\c{c}ats, E-08010 Barcelona, Spain}

\author{A.~M\"oller}
\affiliation{The Research School of Astronomy and Astrophysics, Australian National University, ACT 2601, Australia}
\affiliation{ARC Centre of Excellence for All-sky Astrophysics (CAASTRO), Australia}

\author[0000-0002-2598-0514]{A.~A.~Plazas}
\affiliation{Jet Propulsion Laboratory, California Institute of Technology, 4800 Oak Grove Dr., Pasadena, CA 91109, USA}

\author[0000-0002-9328-879X]{A.~K.~Romer}
\affiliation{Department of Physics and Astronomy, Pevensey Building, University of Sussex, Brighton, BN1 9QH, UK}

\author[0000-0002-9646-8198]{E.~Sanchez}
\affiliation{Centro de Investigaciones Energ\'eticas, Medioambientales y Tecnol\'ogicas (CIEMAT), Madrid, Spain}

\author{V.~Scarpine}
\affiliation{Fermi National Accelerator Laboratory, P. O. Box 500, Batavia, IL 60510, USA}

\author{M.~Schubnell}
\affiliation{Department of Physics, University of Michigan, Ann Arbor, MI 48109, USA}

\author{S.~Serrano}
\affiliation{Institut d'Estudis Espacials de Catalunya (IEEC), 08034 Barcelona, Spain}
\affiliation{Institute of Space Sciences (ICE, CSIC),  Campus UAB, Carrer de Can Magrans, s/n,  08193 Barcelona, Spain}

\author[0000-0002-3321-1432]{M.~Smith}
\affiliation{School of Physics and Astronomy, University of Southampton,  Southampton, SO17 1BJ, UK}

\author{R.~C.~Smith}
\affiliation{Cerro Tololo Inter-American Observatory, National Optical Astronomy Observatory, Casilla 603, La Serena, Chile}

\author[0000-0001-6082-8529]{M.~Soares-Santos}
\affiliation{Brandeis University, Physics Department, 415 South Street, Waltham MA 02453}

\author[0000-0002-7822-0658]{F.~Sobreira}
\affiliation{Laborat\'orio Interinstitucional de e-Astronomia - LIneA, Rua Gal. Jos\'e Cristino 77, Rio de Janeiro, RJ - 20921-400, Brazil}
\affiliation{Instituto de F\'isica Gleb Wataghin, Universidade Estadual de Campinas, 13083-859, Campinas, SP, Brazil}

\author[0000-0002-7047-9358]{E.~Suchyta}
\affiliation{Computer Science and Mathematics Division, Oak Ridge National Laboratory, Oak Ridge, TN 37831}

\author{E.~Swann}
\affiliation{Institute of Cosmology and Gravitation, University of Portsmouth, Portsmouth, PO1 3FX, UK}

\author{M.~E.~C.~Swanson}
\affiliation{National Center for Supercomputing Applications, 1205 West Clark St., Urbana, IL 61801, USA}

\author[0000-0003-1704-0781]{G.~Tarle}
\affiliation{Department of Physics, University of Michigan, Ann Arbor, MI 48109, USA}

\author{B.~E.~Tucker}
\affiliation{ARC Centre of Excellence for All-sky Astrophysics (CAASTRO), Australia}
\affiliation{The Research School of Astronomy and Astrophysics, Australian National University, ACT 2601, Australia}

\author[0000-0001-7211-5729]{D.~L.~Tucker}
\affiliation{Fermi National Accelerator Laboratory, P. O. Box 500, Batavia, IL 60510, USA}

\author{V.~Vikram}
\affiliation{Argonne National Laboratory, 9700 South Cass Avenue, Lemont, IL 60439, USA}

%% Note that the \and command from previous versions of AASTeX is now
%% depreciated in this version as it is no longer necessary. AASTeX 
%% automatically takes care of all commas and "and"s between authors names.

%% AASTeX 6.2 has the new \collaboration and \nocollaboration commands to
%% provide the collaboration status of a group of authors. These commands 
%% can be used either before or after the list of corresponding authors. The
%% argument for \collaboration is the collaboration identifier. Authors are
%% encouraged to surround collaboration identifiers with ()s. The 
%% \nocollaboration command takes no argument and exists to indicate that
%% the nearby authors are not part of surrounding collaborations.

%% Mark off the abstract in the ``abstract'' environment. 
\begin{abstract}

Measurements of the physical properties of accretion disks in active galactic nuclei are important for better understanding the growth and evolution of supermassive black holes. We present the accretion disk sizes of 22 quasars from continuum reverberation mapping with data from the Dark Energy Survey (DES) standard star fields and the supernova C fields. We construct continuum lightcurves with the \textit{griz} photometry that span five seasons of DES observations. These data sample the time variability of the quasars with a cadence as short as one day, which corresponds to a rest frame cadence that is a factor of a few higher than most previous work. We derive time lags between bands with both JAVELIN and the interpolated cross-correlation function method, and fit for accretion disk sizes using the JAVELIN Thin Disk model. These new measurements include disks around black holes with masses as small as $\sim10^7$ $M_{\odot}$, which have equivalent sizes at 2500\AA \, as small as $\sim 0.1$ light days in the rest frame. We find that most objects have accretion disk sizes consistent with the prediction of the standard thin disk model when we take disk variability into account. We have also simulated the expected yield of accretion disk measurements under various observational scenarios for the Large Synoptic Survey Telescope Deep Drilling Fields. We find that the number of disk measurements would increase significantly if the default cadence is changed from three days to two days or one day.

\end{abstract}

%% Keywords should appear after the \end{abstract} command. 
%% See the online documentation for the full list of available subject
%% keywords and the rules for their use.

%% From the front matter, we move on to the body of the paper.
%% Sections are demarcated by \section and \subsection, respectively.
%% Observe the use of the LaTeX \label
%% command after the \subsection to give a symbolic KEY to the
%% subsection for cross-referencing in a \ref command.
%% You can use LaTeX's \ref and \label commands to keep track of
%% cross-references to sections, equations, tables, and figures.
%% That way, if you change the order of any elements, LaTeX will
%% automatically renumber them.
%%
%% We recommend that authors also use the natbib \citep
%% and \citet commands to identify citations.  The citations are
%% tied to the reference list via symbolic KEYs. The KEY corresponds
%% to the KEY in the \bibitem in the reference list below. 

%Introduction
\section{Introduction} \label{sec:intro}

Active galactic nuclei (AGNs) are powered by an accretion disk formed by gas accreted onto a galaxy's central supermassive black hole (SMBH). The accretion disk produces multi-temperature black body emission, with a peak that is typically in the ultraviolet (UV). Studies of the size and structure of the accretion disk are important because they help to understand the growth of SMBHs and the evolution of AGNs.

The conventional accretion disk model is the geometrically thin, optically thick disk model \citep{Shakura1973}. The disk is internally heated by viscous dissipation. Later modifications of the model include external heating by the UV/X-ray source near the SMBH \citep[e.g.][]{Haardt1991}. The disk has a temperature gradient, reaching about $10^5-10^6$ K near the center and getting colder at larger radii \citep[e.g.][]{Shakura1973,Shields1978}. In this model, the temperature profile has the form $T(R) \propto R^{-3/4}$ over a large range of $R$, where $R$ is the distance from the central SMBH. 

It is common to assume that the continuum is well characterized by multi-temperature black body emission where the annulus at radius $R$ is emitting as a black body with temperature $T(R)$ \citep[e.g.][]{Collier1998,Morgan2010,Jiang2017,Mudd17}. We consequently expect longer wavelength emission to primarily originate at larger radii. The disk size at effective wavelength $\lambda$, defined as the position where $kT(R_{\lambda}) = hc/\lambda$, scales with wavelength as $R_{\lambda} \propto \lambda^{\beta}$, where $\beta=4/3$ in the standard thin disk model. Accretion disks are too small to be spatially resolved, and current measurements of the size of accretion disks are mainly from micro-lensing \citep[e.g.][]{Morgan2010} and continuum reverberation mapping \citep[e.g.][]{Shappee2014,Fausnaugh2016,Jiang2017,Mudd17,Homayouni2018}.

Accretion disk size measurements from continuum reverberation mapping rely on measurements of continuum lags between bands at different wavelengths. If the variation of the continuum emission from the accretion disk is driven by the variation of a central illuminating source, such as the ``lamppost'' model \citep[e.g.][]{Cackett2007}, the variation at longer wavelengths is expected to lag the variation at shorter wavelengths due to the light travel time from the inner disk to the outer disk. The lag between two wavelengths $\lambda$ and $\lambda_0$ is
\begin{equation}
\tau = \frac{R_{\lambda_0}}{c}\left[\left(\frac{\lambda}{\lambda_0}\right)^{\beta} - 1 \right]
\label{eq:lag}
\end{equation}
where $R_{\lambda_0}$ is the effective disk size at wavelength $\lambda_0$. Equation (\ref{eq:lag}) states that the disk size is related to the time lag between two lightcurves at different wavelengths. This approach to measuring the accretion disk size is called ``continuum reverberation mapping''. 

One algorithm used to measure lags is the interpolated cross-correlation function (ICCF), which cross-correlates the linearly interpolated lightcurves and calculates the lag as the center or peak of the cross-correlation function \citep[e.g.][]{Peterson1998,Peterson2004}. Another method is JAVELIN, which models the variability of AGNs as a damped random walk (DRW) stochastic process and fits for the lag \citep[e.g.][]{Zu2011,Zu2013}. Simply fitting the continuum lags in different photometric bands with the thin disk model can provide the accretion disk size at a given wavelength. In addition, \citet{Mudd17} presented an alternate method to obtain the disk size, the JAVELIN Thin Disk model, which assumes a thin disk model from the outset and then fits for the thin disk parameters ($R_{\lambda_0}$, $\beta$) directly that best reproduce a series of lightcurves of known effective wavelengths, instead of using the individual lags to find a disk size through Equation (\ref{eq:lag}). 

Early studies to measure accretion disk sizes with continuum reverberation mapping include \citet{Wanders1997} and \citet{Collier1998}, which measured the continuum lags of NGC 7469 at UV and visible wavelengths, respectively. \citet{Sergeev2005} measured the interband lags of 14 AGNs, and found the lags scale with the luminosity as $L^b$ where $b \approx 0.4-0.5$. Recently, several studies obtained accurate measurements of continuum lags using intensive observations spanning from the X-ray to the near-infrared, including \citet{Shappee2014} for NGC 2617, \citet{Edelson2015} and \citet{Fausnaugh2016} for NGC 5548, \citet{Edelson2017} for NGC 4151, \citet{Cackett2018} and \citet{McHardy2018} for NGC 4593 and \citet{Fausnaugh2018} for MCG+08-11-011 and NGC 2617. Other studies have used observations from large sky surveys to measure the accretion disk sizes from larger samples. \citet{Jiang2017} measured the continuum lags of 39 quasars using lightcurves from Pan-STARRS. \citet{Mudd17} used the photometric data on the supernova fields of the Dark Energy Survey (DES) and obtained disk sizes measurements of 15 quasars. \citet{Homayouni2018} presented the continuum lags of 95 quasars from the photometric data for the Sloan Digital Sky Survey (SDSS) Reverberation Mapping (RM) project. 

Some studies, including \citet{Fausnaugh2016} and \citet{Jiang2017}, and most micro-lensing studies like \citet{Morgan2010}, found that the accretion disks are larger than the prediction of the thin disk model by a factor of 2 - 3. One possible explanation of the larger disk sizes is non-thermal disk emission caused by a low density disk atmosphere \citep{Hall2018}. Another explanation is a disk wind that leads to a higher effective temperature in the outer part of the accretion disk \citep[e.g.][]{Sun2018,Li2018}. \citet{Gaskell2017} found the internal reddening of AGNs that leads to an underestimation of the far-UV luminosity can be an explanation of the discrepancy as well. Contamination by the diffuse continuum from the broad line region (BLR) \citep[e.g.][]{Korista2001,Lawther2018} may lead to larger disk size measurements from continuum reverberation mapping. In addition, an inhomogeneous disk with significant local temperature fluctuations may also explain the larger disk sizes from micro-lensing studies \citep{Dexter2011}. However, \citet{Mudd17} and \citet{Homayouni2018} did not find systematic trends of larger disk sizes than the prediction of the thin disk model. To address this discrepancy, more disk size measurements are required, especially using high-cadence time series data.

In this paper, we present disk size measurements using the photometric data in the DES standard star fields and the supernova C (SN-C) fields. The structure of this paper is as follows. Section \ref{sec:obs} introduces the photometric data we use. Section \ref{sec:analysis} introduces the methodology and results of the time series analysis, including the lag and disk size measurements. In Section \ref{sec:verification} 
we discuss various tests we performed to verify the measurements. In Section \ref{sec:bhmass_disksize} we describe our measurements of the correlation between the accretion disk size and the mass of the SMBH. Section \ref{sec:nolagobj} discusses the objects without lag measurements. In Section \ref{sec:lsst} we present simulations of the Large Synoptic Survey Telescope Deep Drilling Fields and quantify the effect of observational cadence on lag measurements. Section \ref{sec:summary} summarizes the paper. Throughout the paper we adopt $\Lambda$CDM cosmology with $H_0 = 70 \, {\rm km/s/Mpc}$, $\Omega_m = 0.3$, $\Omega_{\Lambda} = 0.7$.

%Observations
\section{Observations} \label{sec:obs}

DES is a ground-based, wide-area, visible and the near-infrared imaging survey \citep{DESDR1}. DES started Commissioning and Science Verification (SV) observations in 2012, and the main survey began in 2013. DES uses the Dark Energy Survey Camera (DECam), a 570 megapixel, $2.2^{\circ}$ field of view camera installed on the 4-m Victor M. Blanco telescope at the Cerro Tololo Inter-American Observatory \citep{DECam}. DES includes a $5000 \, {\rm deg}^2$ wide-area survey in the \textit{grizY} bands, and $27 \, {\rm deg}^2$ in the \textit{griz} bands that are repeatedly imaged to identify and characterize supernova. DES typically observes standard star fields in morning and evening twilight for calibration, and occasionally around midnight as well. Standard star observations can have a nightly or even higher observational cadence in some fields, with a typical exposure time of 15 seconds for a single epoch. The high observational cadence supports accurate photometric RM analysis, despite the short exposure time. 

We use standard star observations from the DES SV period through Year 4 (Y4) in the MaxVis field, the C26202 fields, and 6 other fields within the SDSS footprint. We incorporate spectroscopic data for the MaxVis and the C26202 fields from the Australian DES/Optical redshifts for DES (OzDES) program, a spectroscopic survey with the Anglo-Australian Telescope that was designed to follow up targets identified from DES \citep{OzDES,Childress2017}.

%Observations-MaxVis Field
\subsection{MaxVis Field} \label{subsec:obs-maxvis}

\begin{figure}%[ht!]
%\figurenum{1}
\plotone{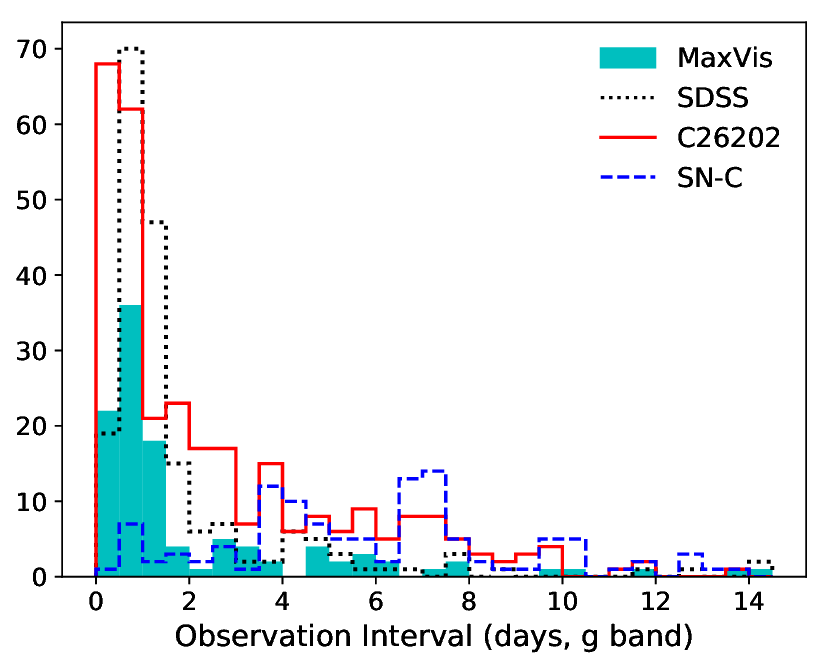}
\figcaption{Time interval distribution between consecutive pairs of epochs for the lightcurve of a representative object in each field. The cyan filled, black dotted, red solid and blue dashed histograms are the cadence distribution of DES J063037.48$-$575610.30 in the MaxVis field, DES J005905.51+000651.66 in the SDSS fields, DES J033408.25$-$274337.81 in the C26202 field and DES J034001.53$-$274036.91 in the SN-C fields, respectively.\label{fig:obscad}}
\end{figure}

\begin{figure}%[ht!]
%\figurenum{1}
\plotone{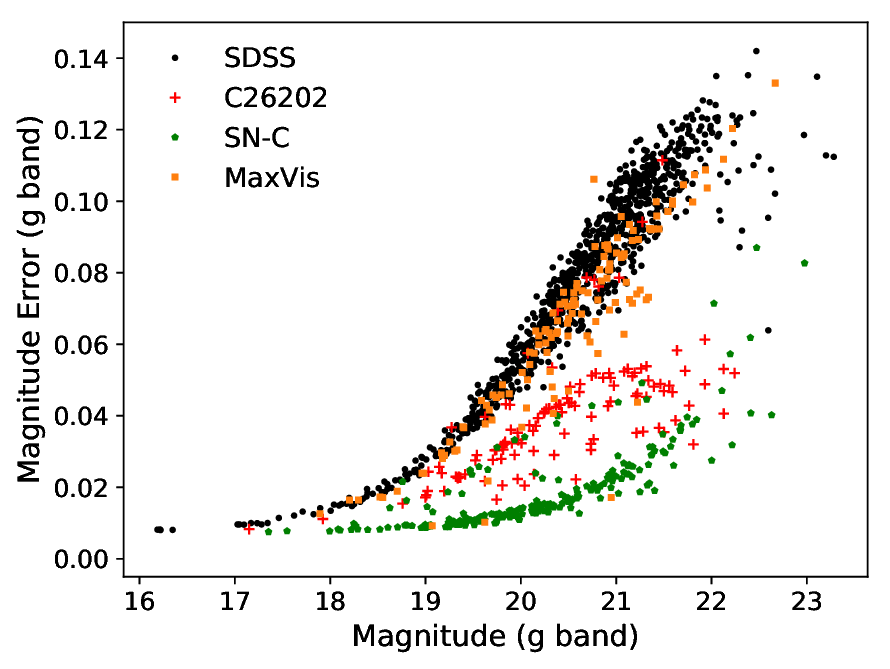}
\figcaption{Magnitude uncertainty as a function of magnitude for \textit{g}-band data for the four datasets. The orange squares, black circles, red crosses and green pentagons represent the objects in the MaxVis, the SDSS, the C26202 and the SN-C fields, respectively. The magnitude shown here is calculated as the mean magnitude of the object from SV to Y4, and the magnitude uncertainty is calculated as the mean magnitude uncertainty during this period. The magnitude uncertainties have included the calibration errors.\label{fig:obsdepth}}
\end{figure}

The MaxVis field is centered at $RA= 97.5^{\circ}$, $DEC=-58.75^{\circ}$. The field is observable throughout the DES observing season, and it was named because of this ``Maximum Visibility''. We visually inspected all OzDES spectra flagged as non-stellar within the field, and selected 130 quasars as candidates with spectroscopic redshifts from OzDES Global Redshift Catalog \citep{Childress2017}. The cyan histogram in Figure \ref{fig:obscad} shows the observational cadence distribution of DES J063037.48$-$575610.30, a representative object in the MaxVis field. The distribution peaks around one day, indicating that most epochs for this object are obtained with a nearly daily cadence. The orange squares in Figure \ref{fig:obsdepth} shows the magnitude uncertainty as a function of the \textit{g}-band magnitude for the quasars in the MaxVis field. The depth of the MaxVis field is intermediate relative to the other standard star fields.

%Observations-SDSS
\subsection{SDSS Stripe 82 Fields} \label{subsec:obs-sdss}
We use six fields near the celestial equator that overlap the SDSS Stripe 82 field \citep[e.g.][]{Adelman2007}. We match DES observations to the color and variability selected quasar catalog from \citet{Peters2015}, and select 884 objects as candidates with more than 200 epochs in the \textit{grizY} bands from SV to Y4, among which 593 objects have spectroscopic redshifts. Similar to the MaxVis field, Figure \ref{fig:obscad} shows that the observational cadence in the SDSS fields is roughly daily. On the other hand, Figure \ref{fig:obsdepth} shows that the observations in the SDSS fields are shallower compared to the other fields.

%Observations-C26202
\subsection{C26202 and SN-C Fields} \label{subsec:obs-c26202}
The C26202 field is centered on the standard star C26202 at $RA=53.1^{\circ}$, $DEC=-27.85^{\circ}$, which overlaps with the DES C1, C2 and C3 supernova fields. We match the DES detections within a circle of $3.5^{\circ}$ radius to the spectroscopically confirmed quasar catalog presented by \citet{Tie2017}, and selected 318 quasars as candidates with more than 200 epochs in the \textit{griz} bands from SV to Y4. Note that the candidates include not only quasars from the C26202 standard star field, but also quasars within the SN-C fields. The DES observations of the C26202 standard star field include both approximately daily standard star observations and approximately weekly supernova observations. Figure \ref{fig:obscad} shows the cadence distribution of a representative object in this field, where most observations are high-cadence standard star observations shown by the peak around one day, while there are also supernova observations with longer intervals between epochs. Figure \ref{fig:obsdepth} shows that quasars within the C26202 standard star field have the deepest observations (the red crosses) compared to the quasars within the other two, while the observations of quasars within the supernova fields (the green pentagons) are even deeper. Hereafter we use ``C26202 fields'' to refer to both the C26202 standard star field and the SN-C fields unless otherwise specified.

%Time Series Analysis
\section{Time Series Analysis} \label{sec:analysis}

We construct lightcurves using photometric data from the DES Y4A1 catalogs. We adopt the PSF magnitude and its error for the photometry, and exclude bad epochs based on the DES data quality flags. For time series analysis, the calibration between epochs would add additional systematic errors. We adopt the typical error of the DES Forward Global Calibration Method (FGCM) from \citet{Burke2018} and combine it with the magnitude errors by quadrature to calculate the total uncertainty of each single epoch. 

There are epochs separated by only minutes, which can be caused by the short acquisition images or the cosmic ray separation of the supernova observations. Since the variations between these epochs are not likely to be intrinsic to the quasars, we exclude the short acquisition epochs and combine the supernova epochs separated by less than half hour to avoid possible artifacts in further analysis. We assume that the seeing and the sky transparency do not vary significantly within this time range, and calculate the magnitude and magnitude error of the combined epoch as 
\begin{equation}
m_{comb} = \frac{\sum \limits_i m_i/\sigma_i^2}{\sum \limits_i 1/\sigma_i^2} \, , \quad \sigma_{comb}^2 = \frac{1}{\sum \limits_i 1/\sigma_i^2}
\end{equation}
where $m_i$ and $\sigma_i$ are the magnitude and its statistical error of each single epoch. We combine $\sigma_{comb}$ with the calibration error by quadrature to calculate the total uncertainty of the combined epoch.  

We assess the variability of an object by calculating a $\chi^2$ value defined as
\begin{equation}
\chi^2 = \sum \limits_X \sum \limits_i \left (\frac{m_{i,X}-\overline{m}_X}{\sigma_{i,X}} \right )^2 
\label{eq:chi2}
\end{equation}
where $m_{i,X}$ and $\sigma_{i,X}$ represents the magnitude and uncertainty of the \textit{i}th data point in band X (X=\textit{g},\textit{r},\textit{i},\textit{z},\textit{Y}), and $\overline{m}_X$ represents the mean magnitude in band X. We calculate the $\chi^2$ values both for the whole SV - Y4 period and for the single seasons. We perform a time series analysis on objects and seasons that satisfy: (1) $N_{fit}>100$, where $N_{fit}$ is the number of epochs in the season to analyze; (2) $N_{other}>200$, where $N_{other}$ is the number of epochs in all the other seasons that are not analyzed; and (3) $\chi_r^2 > 2$, indicating significant variability, where $\chi_r^2 = \chi^2/(N_{fit}-5)$ is the reduced $\chi^2$. There are 48 objects in the MaxVis field, 457 objects in the SDSS fields and 297 objects in the C26202 fields that met the criteria in at least one observational season. Among these objects, about half of the objects in the MaxVis field and the SDSS fields have only one season that met the criteria, while most objects in the C26202 fields have at least two seasons. We analyze each season that passed the selection independently for computational convenience when measuring time lags. Quasars can have lag detections from multiple seasons, and we discuss this further in Section \ref{subsec:anl-javelin}. 

\begin{figure}%[ht!]
%\figurenum{1}
\plotone{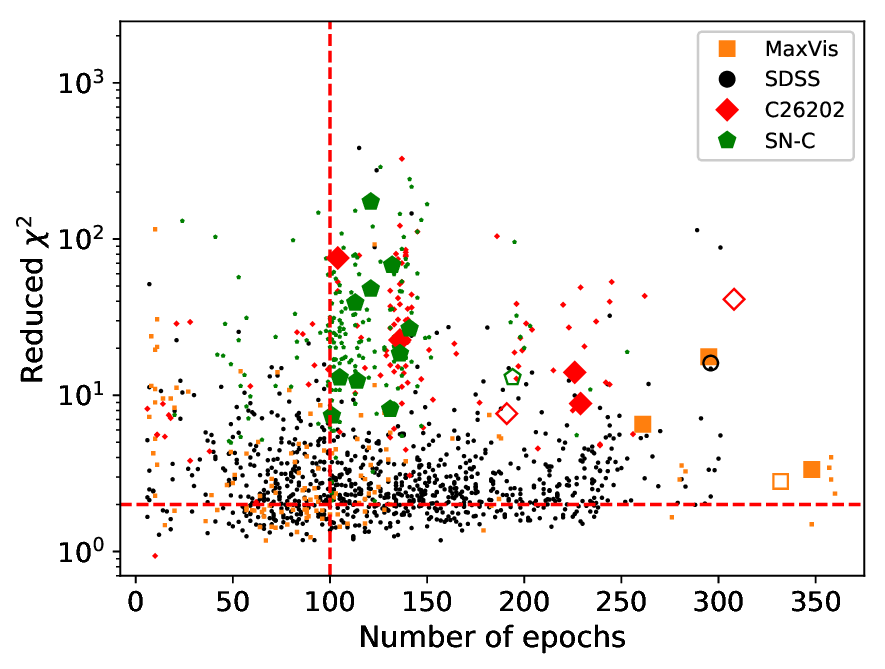}
\figcaption{$\chi_r^2$ as a function of the number of epochs. The orange squares, black circles, red diamonds and green pentagons represent the objects from the MaxVis field, the SDSS fields, the C26202 standard star field, and the SN-C fields, respectively. The small symbols represent the objects where we do not obtain good lag measurements. The large symbols represent the objects in the main sample with the empty symbols for the flagged objects. For the objects without lag measurements, $\chi_r^2$ and the number of epochs are calculated from the season where $\chi_r^2$ is the largest. For objects in the main sample, $\chi_r^2$ and the number of epochs are from the season where the lags and disk sizes are measured. If an object has lag measurements from multiple seasons, $\chi_r^2$ is from the earliest season that has lag measurements. The horizontal red dashed line is drawn at $\chi_r^2=2$, and the vertical line is drawn at $N_{fit}=100$. \label{fig:obs_chi2_nexp}}
\end{figure}

Figure \ref{fig:obs_chi2_nexp} shows $\chi_r^2$ as a function of the number of epochs for the candidates. The quasars within the SDSS fields show the smallest variations relative to the photometric errors, while many quasars in the SN-C field have large $\chi_r^2$ due to the deep supernova observations.

In this work, we use JAVELIN, the JAVELIN Thin Disk model, and the ICCF method to measure lags and derive disk sizes. We obtain good quality disk size measurements for 22 quasars. We refer to these quasars as the ``main sample''. The basic properties of these quasars are listed in Table \ref{tab:objinfo}, and disk size results are presented in Table \ref{tab:objresult}. Seventeen of the 22 quasars form the most reliable subset of the measurements. The remaining five are flagged due to inconsistencies of lags either between the observational seasons ($flag=1$) or the analysis methods ($flag=2$), or simulation results that imply lower reliability based on the observational data and an estimate of the disk size ($flag=3$). The process of assigning flags is described in the next sections. We separate objects with and without flags in Table \ref{tab:objresult}. 

\begin{figure*}%[ht!]
%\figurenum{1}
\gridline{\fig{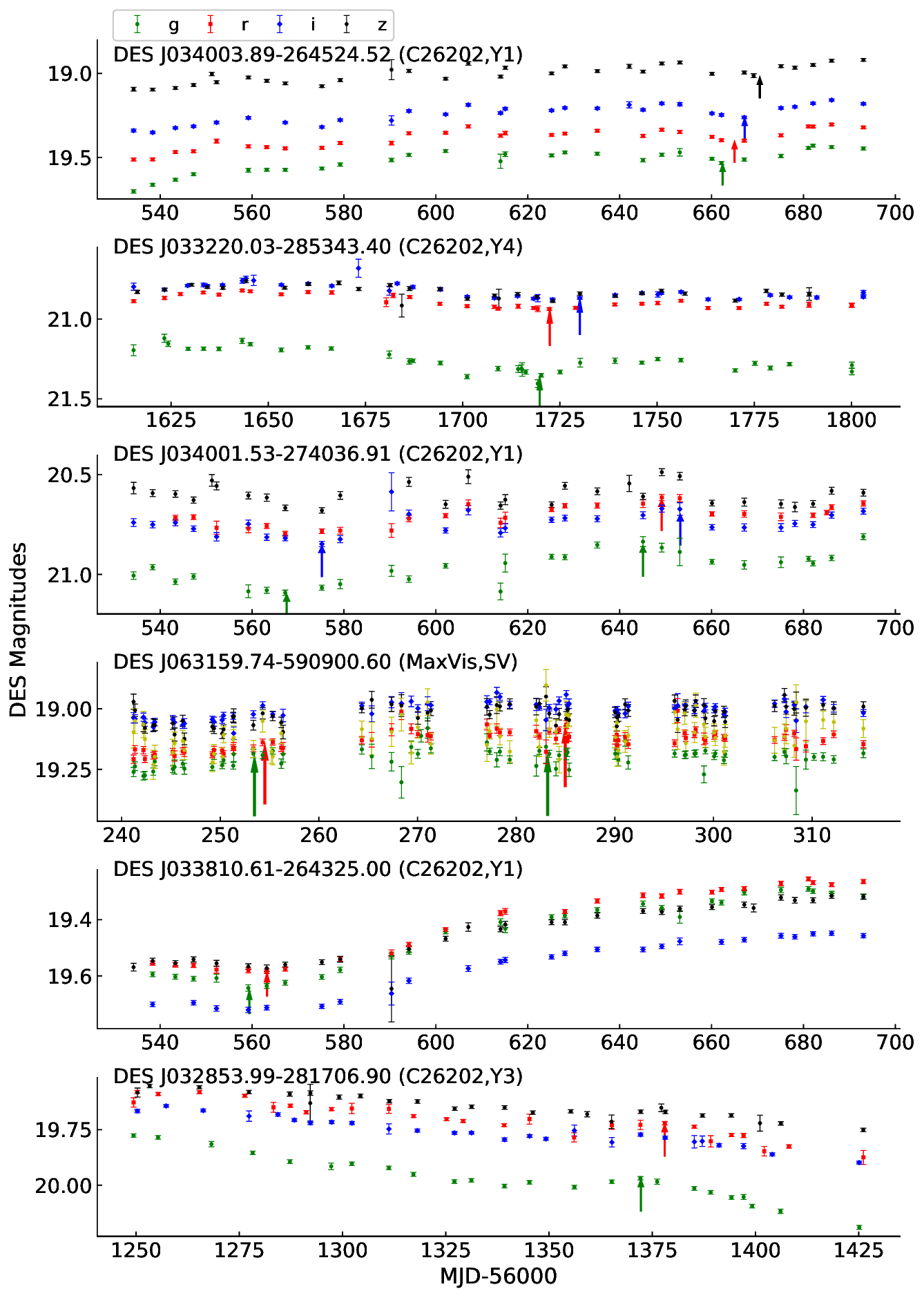}{0.88\textwidth}{}}
%\epsscale{2.0}
\figcaption{DES lightcurves of a randomly selected subset of the unflagged objects in the main sample. The object's name, field and the observational season(s) from which we measure the lag is shown in the upper left corner of each row. The green circles, red squares, blue diamonds, black hexagons and yellow pentagons represent the \textit{g}, \textit{r}, \textit{i}, \textit{z} and \textit{Y} data, respectively. The green, red, blue and black arrows point to the approximate positions of the features that show visible lags in the \textit{g}, \textit{r}, \textit{i} and \textit{z}-band lightcurves, respectively.\label{fig:lcexamp}}
\end{figure*}

Figure \ref{fig:lcexamp} shows the lightcurves of six randomly selected unflagged objects in the observational season from which we obtain lag measurements. We present the lightcurve data and plots in the online journal for all seasons and objects in the main sample.

%\begin{figure*}%[ht!]
%\figurenum{1}
%\gridline{\fig{LC/lcgood_Y4add_p2.eps}{1.0\textwidth}{}}
%\epsscale{2.0}
%\figcaption{Figure \ref{fig:lcgood1}, continued\label{fig:lcgood2}}
%\end{figure*}

%\begin{figure*}%[ht!]
%\figurenum{1}
%\gridline{\fig{LC/lcgood_Y4add_p3.eps}{1.0\textwidth}{}}
%\epsscale{2.0}
%\figcaption{Figure \ref{fig:lcgood1}, continued\label{fig:lcgood3}}
%\end{figure*}

\begin{deluxetable}{ccccc}
\tablecaption{Quasars in the Main Sample\label{tab:objinfo}}
%\tablewidth{5pt}
\tabletypesize{\tiny}
\tablehead{
\colhead{Object Name} & \colhead{z} & \colhead{Field} & \colhead{Reference Line} & \colhead{$M_{BH}$} \\
 &  &  &  & \colhead{($10^8 M_{\odot}$)}
}
%\decimalcolnumbers
\startdata
DES J063037.48$-$575610.30 & 0.43 & MaxVis & H$\beta$ & 1.23 \\
DES J063510.91$-$585303.70 & 0.22 & MaxVis & H$\beta$ & 0.23 \\
DES J063159.74$-$590900.60 & 0.73 & MaxVis & Mg II & 2.00 \\
DES J062758.99$-$582929.60 & 0.49 & MaxVis & H$\beta$ & 0.20 \\
DES J033002.93$-$273248.30 & 0.53 & C26202 & H$\beta$ & 0.80 \\
DES J033408.25$-$274337.81 & 1.03 & C26202 & Mg II & 1.18 \\
DES J034003.89$-$264524.52 & 0.49 & C26202 & H$\beta$ & 0.44 \\
DES J032724.94$-$274202.81 & 0.76 & C26202 & Mg II & 1.71 \\
DES J033545.58$-$293216.51 & 0.72 & C26202 & Mg II & 0.82 \\
DES J033810.61$-$264325.00 & 0.85 & C26202 & Mg II & 2.05 \\
DES J034001.53$-$274036.91 & 1.15 & C26202 & Mg II & 1.98 \\
DES J033853.20$-$261454.82 & 1.17 & C26202 & Mg II & 2.65 \\
DES J033051.45$-$271254.90 & 0.63 & C26202 & H$\beta$ & 0.61 \\
DES J032853.99$-$281706.90 & 1.00 & C26202 & Mg II & 5.65 \\
DES J032801.84$-$273815.72 & 1.59 & C26202 & Mg II & 3.28 \\
DES J033230.63$-$284750.39 & 0.86 & C26202 & Mg II & 1.49 \\
DES J033729.20$-$294917.51 & 1.35 & C26202 & Mg II & 1.95 \\
DES J033220.03$-$285343.40 & 1.27 & C26202 & Mg II & 1.37 \\
DES J033342.30$-$285955.72 & 0.55 & C26202 & H$\beta$ & 2.05 \\
DES J033052.19$-$274926.80 & 1.95 & C26202 & C IV & 0.98 \\
DES J033238.11$-$273945.11 & 0.84 & C26202 & Mg II & 1.33 \\
DES J005905.51+000651.66 & 0.72 & SDSS & H$\beta$ & 8.87
\enddata
\tablecomments{Basic parameters of the DES quasars in the main sample. Column (1) gives the names of the objects. Column (2) gives the redshifts. Column (3) gives the field names. Column (4) gives the emission lines used to estimate to black hole mass, and Column (5) gives the single-epoch estimate of the black hole mass (See Section \ref{sec:bhmass_disksize}). The uncertainty of the black hole mass is about 0.4 dex.}
\end{deluxetable}

%Analysis-JAVELIN
\subsection{JAVELIN Analysis} \label{subsec:anl-javelin}

We use JAVELIN \citep{Zu2011} to model quasar variability as a DRW. The covariance function of a DRW has an exponential form
\begin{equation}
S(\Delta t) = \sigma_{\mbox{\tiny DRW}}^2 \, {\rm exp}(-|\Delta t / \tau_{\mbox{\tiny DRW}}|)
\label{eq:cov_drw}
\end{equation}
where $\Delta t$ is the time interval between two epochs, $\sigma_{\mbox{\tiny DRW}}$ is the amplitude, and $\tau_{\mbox{\tiny DRW}}$ is the characteristic time scale. Previous studies have shown that quasar lightcurves are generally well described by a DRW \citep[e.g.][]{Kelly2009,Kozlowski2010,MacLeod2010,Zu2013}. However, several studies found that the Kepler lightcurves of AGNs show steeper power spectral density (PSD) than the DRW model at time scales shorter than $\sim$ month \citep[e.g.][]{Mushotzky2011,Smith2018}. Here we use JAVELIN with the awareness of the potential effect of this deviation from DRW. 

JAVELIN first fits the continuum lightcurve, which can be the lightcurve in any of the broad bands in continuum RM, to constrain $\sigma_{\mbox{\tiny DRW}}$ and $\tau_{\mbox{\tiny DRW}}$. Then JAVELIN assumes that the line lightcurve, which in our case is the lightcurve in another continuum band, is a shifted, smoothed and scaled version of the first lightcurve. This fits three additional parameters: the time lag, the top-hat smoothing factor and the flux scaling factor. We fix $\tau_{\mbox{\tiny DRW}}$ to the value from the \textit{g}-band continuum fitting when fitting for the time lag of most objects, since in the case of an accretion disk the time lag is much smaller than the time scale of the DRW and the fitting result is insensitive to $\tau_{\mbox{\tiny DRW}}$. 

\begin{figure*}%[ht!]
%\figurenum{1}
\gridline{\fig{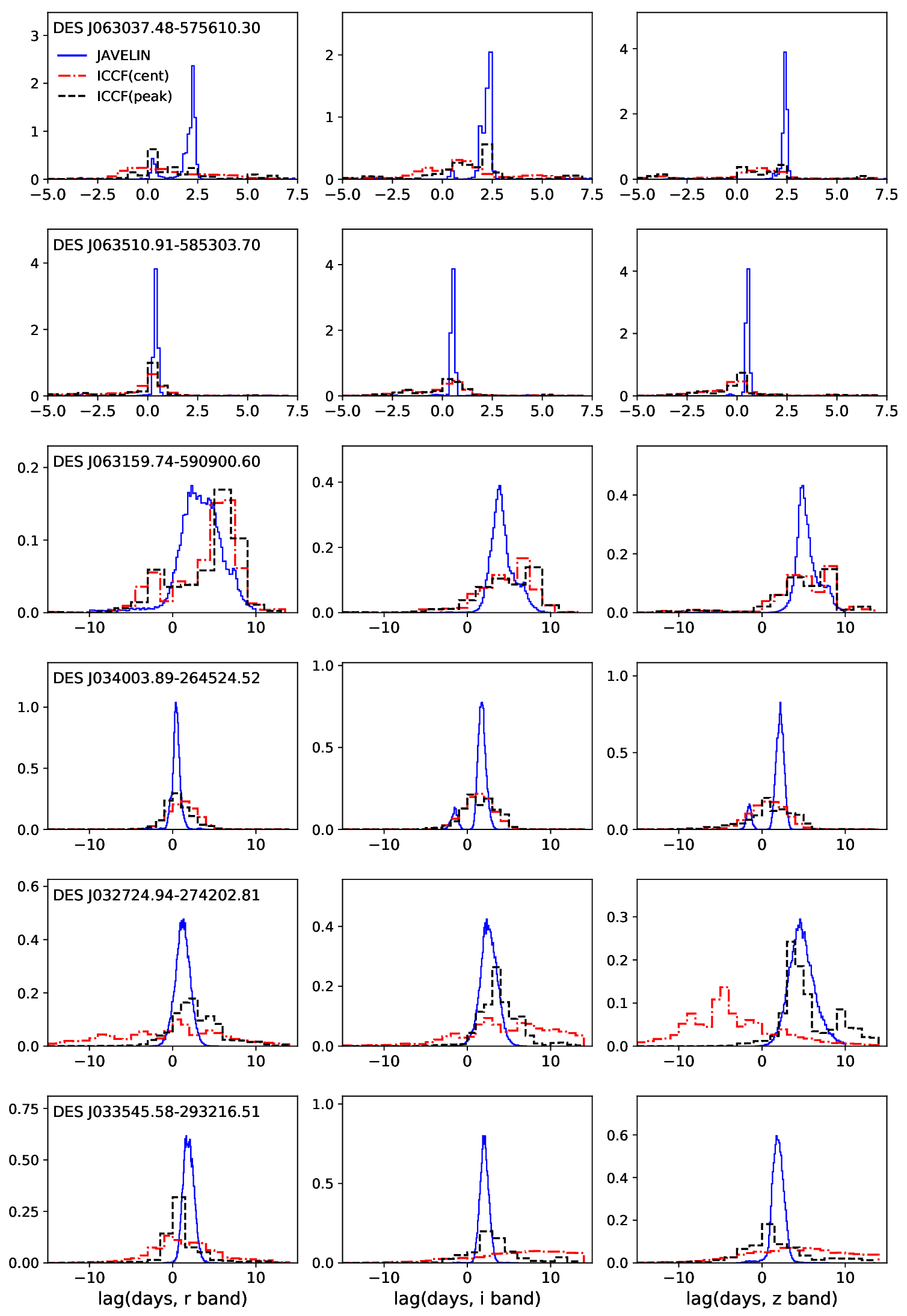}{0.8\textwidth}{}}
%\epsscale{2.0}
\figcaption{Probability distributions of time lags for quasars without flags in the main sample. Each row represents the results for one object whose name is listed in the upper left corner of the first panel, with the first, second and third columns representing the lags in the \textit{r}, \textit{i} and \textit{z} band relative to \textit{g} band, respectively. In each panel, the blue solid line is the lag distribution from JAVELIN, while the red dash-dotted line and the black dashed line represent the ICCF center and peak distribution, respectively.\label{fig:lagclean1}}
\end{figure*}

\begin{figure*}%[ht!]
%\figurenum{1}
\gridline{\fig{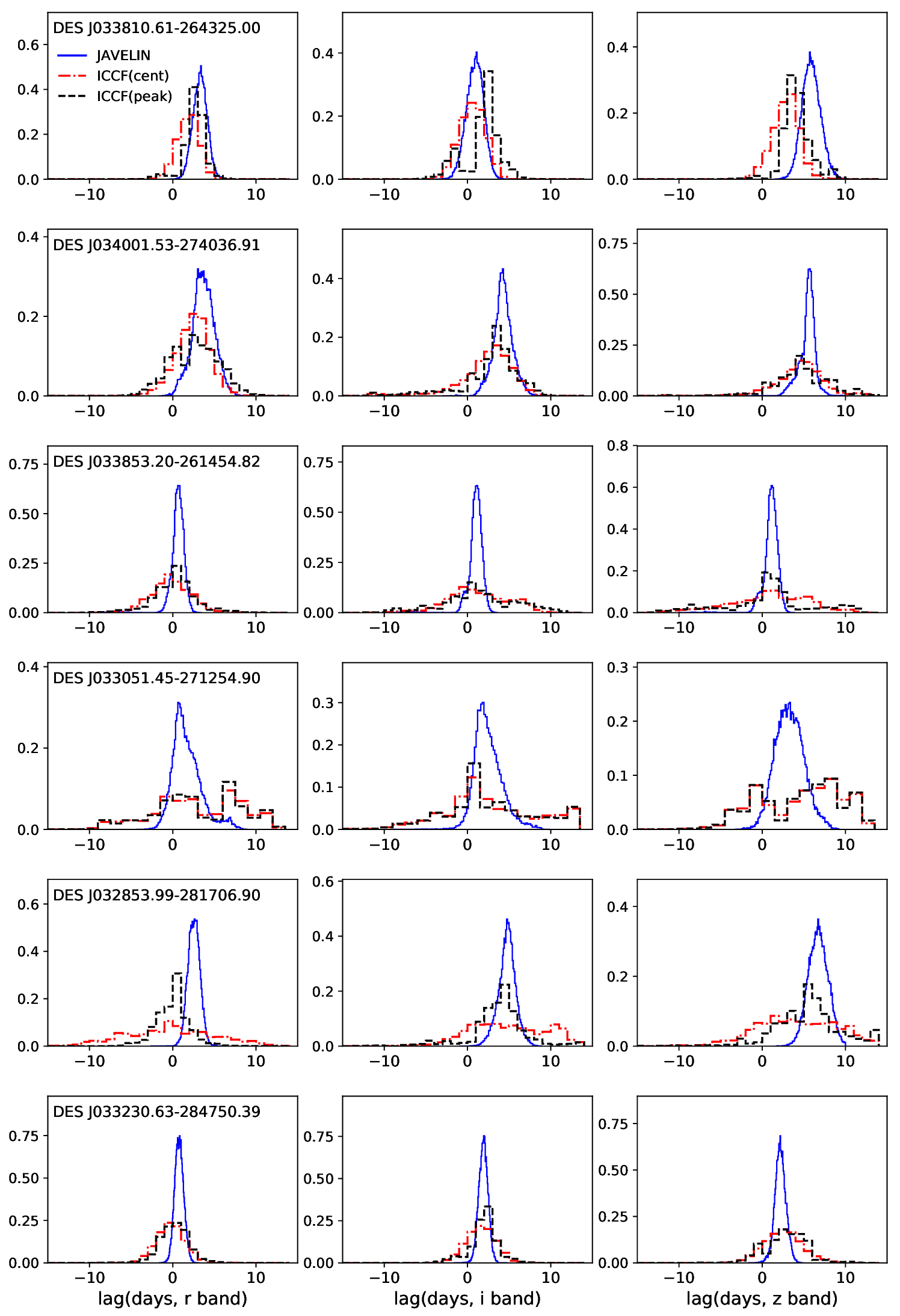}{0.8\textwidth}{}}
%\epsscale{2.0}
\figcaption{Figure \ref{fig:lagclean1}, continued.\label{fig:lagclean2}}
\end{figure*}

\begin{figure*}%[ht!]
%\figurenum{1}
\gridline{\fig{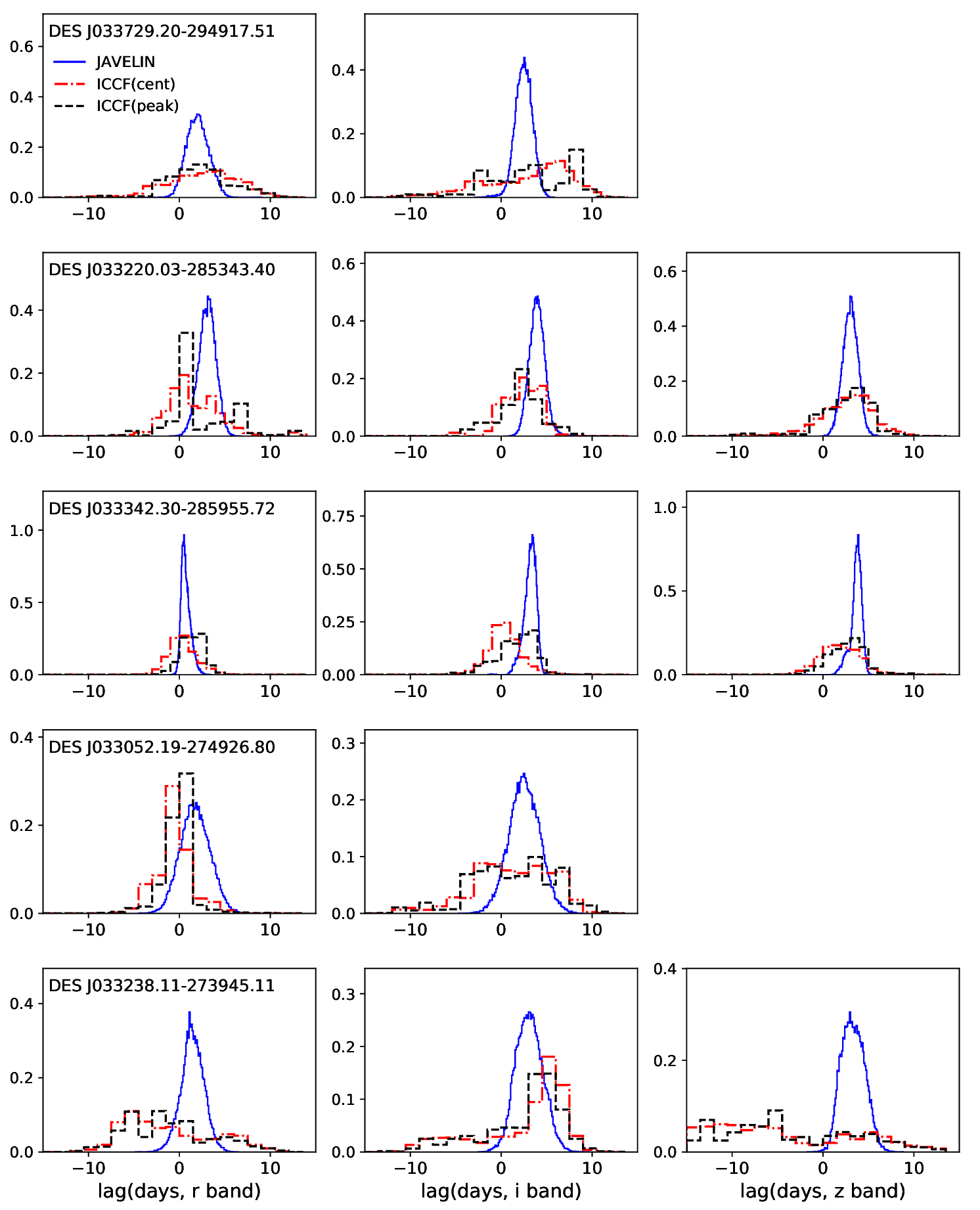}{0.8\textwidth}{}}
%\epsscale{2.0}
\figcaption{Figure \ref{fig:lagclean1}, continued.\label{fig:lagclean3}}
\end{figure*}

\begin{figure*}%[ht!]
%\figurenum{1}
\gridline{\fig{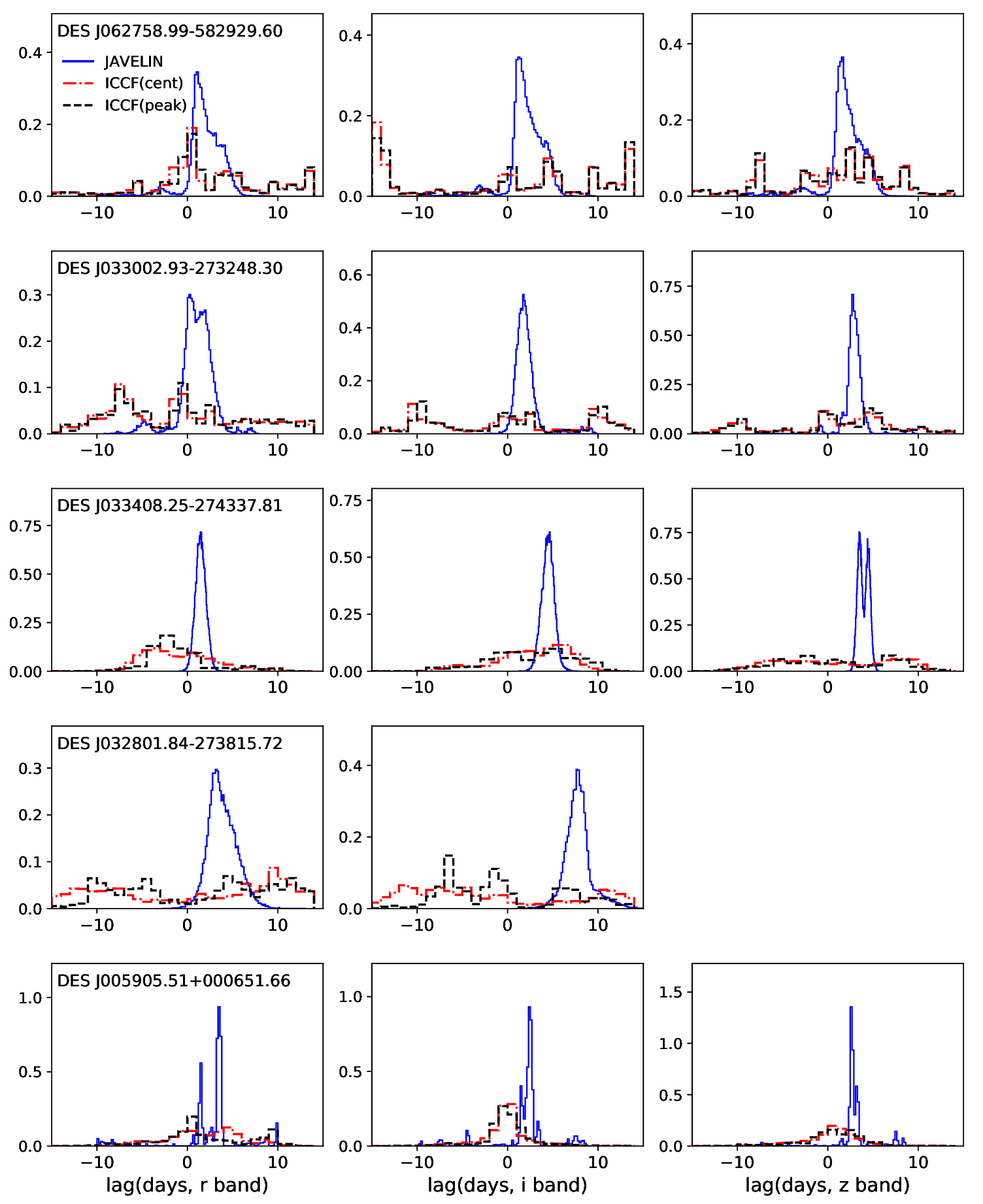}{0.8\textwidth}{}}
%\epsscale{2.0}
\figcaption{Same as Figures \ref{fig:lagclean1} - \ref{fig:lagclean3}, but for quasars with flags in the main sample. \label{fig:lagflag1}}
\end{figure*}

Figures \ref{fig:lagclean1} - \ref{fig:lagflag1} show the probability distribution of time lags of the \textit{r}, \textit{i} and \textit{z} bands relative to the \textit{g} band. Figures \ref{fig:lagclean1} - \ref{fig:lagclean3} only include objects without flags, while Figure \ref{fig:lagflag1} shows flagged objects. In most cases there is a single, clear peak in the lag distribution. While the distributions show secondary peaks in some objects, the amplitude of the secondary peak is very small compared to the main peak. Objects from the MaxVis field, whose observational cadence is around 1-day, produce lags with significantly smaller uncertainty compared to some of the objects in the C26202 field with about a 7-day cadence. We adopt the median of the probability distribution as the best-fit lag, and the 16th and 84th percentile of the distribution as the 1$\sigma$ lower and upper limit of the lag, respectively. We provide more detailed comments on individual objects in the Appendix.

\begin{figure*}%[ht!]
%\figurenum{1}
\gridline{\fig{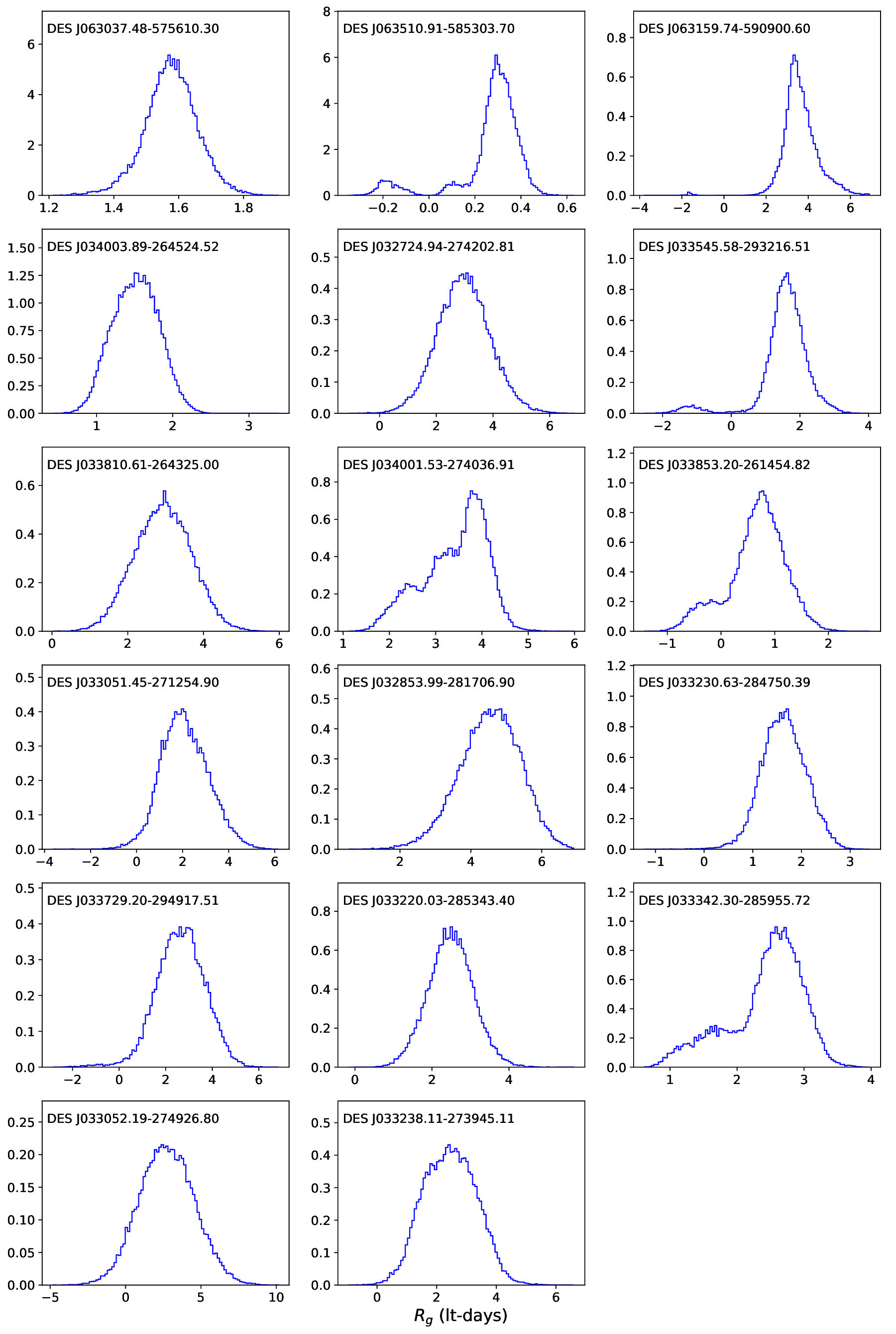}{0.8\textwidth}{}}
%\epsscale{2.0}
\figcaption{Probability distributions for the observed frame \textit{g}-band accretion disk sizes from the JAVELIN Thin Disk model for quasars without flags in the main sample. Each panel shows the result for one object.\label{fig:alphagood_clean}}
\end{figure*}

\begin{figure*}%[ht!]
%\figurenum{1}
\gridline{\fig{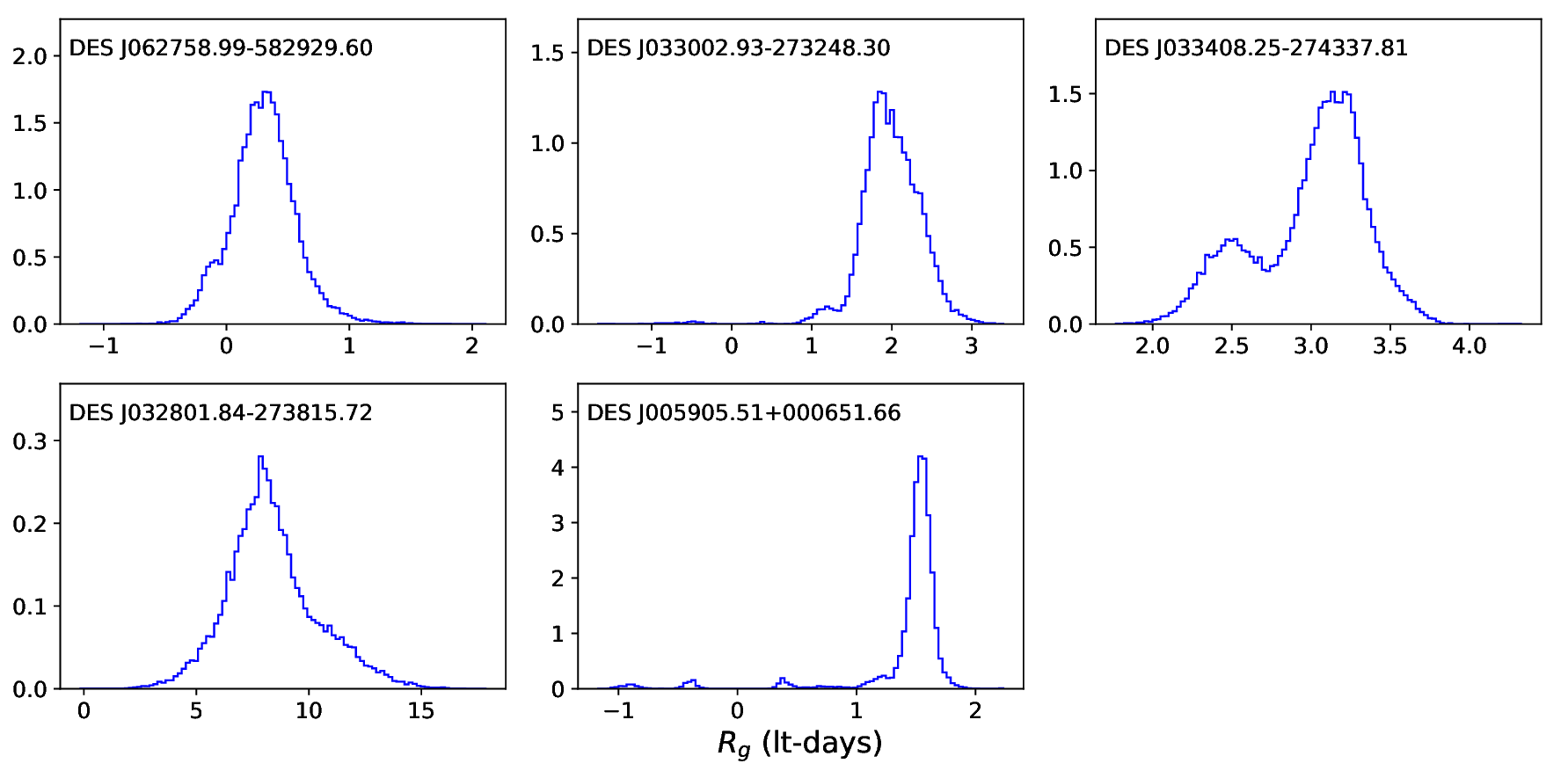}{1.0\textwidth}{}}
%\epsscale{2.0}
\figcaption{Same as Figure \ref{fig:alphagood_clean}, but for quasars with flags in the main sample. \label{fig:alphagood_flag}}
\end{figure*}

We also use the JAVELIN Thin Disk model extension developed by \citet{Mudd17} to measure the accretion disk size $R_{\lambda_0}$ and index $\beta$ in Equation (\ref{eq:lag}) using all four bands simultaneously. The JAVELIN Thin Disk model makes use of the information from photometric lightcurves in all bands to better constrain the accretion disk size, and reduces the number of parameters. Similar to \citet{Mudd17}, we find that the JAVELIN Thin Disk model could not well-constrain both $R_{\lambda_0}$ and $\beta$ at the same time, so we fix $\beta=4/3$ during the fitting. We again fix $\tau_{\mbox{\tiny DRW}}$. Figure \ref{fig:alphagood_clean} and Figure \ref{fig:alphagood_flag} show the probability distribution of the \textit{g}-band accretion disk size $R_g$ in the observed frame, i.e. the disk size at $\lambda = 4730 (1+z) \, {\rm \AA}$, where $z$ is the redshift. Figure \ref{fig:alphagood_clean} only includes objects without flags, while Figure \ref{fig:alphagood_flag} shows flagged objects. Again, most distributions show clear single peaks. We convert the \textit{g}-band disk size to rest frame 2500 \AA\, assuming $R_{\lambda} \propto \lambda^{4/3}$ for comparison with other studies. 

We inspected the lag distributions from JAVELIN and the disk size distributions from the JAVELIN Thin Disk model for all candidates that satisfied the criteria in Section \ref{sec:analysis}. We define a successful measurement of the single-band lag or the accretion disk size if the probability distributions satisfy three criteria: (1) The 1$\sigma$ lower limit of the lag or disk size is larger than zero; (2) The probability distribution shows a clear single peak without significant secondary peaks. We define a secondary peak to be ``significant'' if its peak probability density is more than 20\% of the main peak. Secondary peaks may also be a ``bump'' in the main peak. We treat a ``bump'' as a secondary peak if it is separated from the main peak by at least 0.75 days, and the probability density difference between the bump peak and where it connects the main peak is larger than 7.5\% of the main peak. (3) The 1$\sigma$ upper limit of the top-hat smoothing factor is smaller than 30. We added the third criterion because large smoothing factors are usually related to smooth lag distributions without clear peaks, which is a common feature of the failed fits from JAVELIN. We define successful measurements for an \textit{object} in an observational season if the lag distributions in at least one of the \textit{riz} bands and the disk size distributions satisfy the successful measurement criteria. We identify 22 quasars to have successful lag and disk size measurements in at least one observational season, and we refer to these quasars as the ``main sample''. Table \ref{tab:objinfo} lists some of the basic parameters of the quasars in the main sample. Table \ref{tab:objresult} shows the best-fit lag and the 1$\sigma$ errors from JAVELIN in Columns (2)-(4), and the 2500\AA \, accretion disk size in Column (5). While the selection criteria requires positive lags in only one band, nearly all of the quasars in the main sample have positive lags in at least two of the \textit{riz} bands relative to the \textit{g} band.

%\begin{figure*}%[ht!]
%\figurenum{1}
%\gridline{\fig{Jev/laggood_flag_p2.eps}{0.8\textwidth}{}}
%\epsscale{2.0}
%\figcaption{Figure \ref{fig:lagflag1}, continued.\label{fig:lagflag2}}
%\end{figure*}

\begin{deluxetable}{lCCCCcc}[ht!]
\tablecaption{Lags and Accretion Disk Sizes\label{tab:objresult}}
%\tablewidth{5pt}
\tabletypesize{\tiny}
\tablehead{
\colhead{Object Name(Season)} & \colhead{$\tau_r$} & \colhead{$\tau_i$} & \colhead{$\tau_z$} & \colhead{$R_{2500\AA}$} & \colhead{flag} & \colhead{Visible}\\
 & \colhead{(days)} & \colhead{(days)} & \colhead{(days)} & \colhead{(lt-days)} &  &\colhead{lag}
}
%\decimalcolnumbers
\startdata
DES J0630$-$5756(SV) & 2.2^{+0.2}_{-0.9} & 2.3^{+0.1}_{-0.4} & 2.4^{+0.1}_{-0.1} & 0.76^{+0.04}_{-0.04} & 0 & Y \\
DES J0635$-$5853(SV) & 0.4^{+0.1}_{-0.1} & 0.5^{+0.1}_{-0.1} & 0.5^{+0.1}_{-0.1} & 0.14^{+0.03}_{-0.04} & 0 & N \\
DES J0631$-$5909(SV) & 3.4^{+2.4}_{-2.2} & 4.0^{+1.7}_{-1.0} & 5.2^{+1.7}_{-0.9} & 1.81^{+0.41}_{-0.28} & 0 & Y \\
DES J0340$-$2645(Y1) & 0.4^{+0.4}_{-0.4} & 1.7^{+0.5}_{-0.6} & 2.1^{+0.5}_{-0.7} & 0.74^{+0.15}_{-0.16} & 0 & Y \\
DES J0327$-$2742(Y3,4) & 1.2^{+0.9}_{-0.9} & 2.6^{+1.0}_{-0.9} & 4.7^{+1.6}_{-1.4} & 1.53^{+0.47}_{-0.46} & 0 & Y \\
DES J0335$-$2932(Y1,3) & 1.9^{+0.7}_{-0.6} & 2.0^{+0.5}_{-0.5} & 1.9^{+0.7}_{-0.6} & 0.83^{+0.25}_{-0.23} & 0 & Y \\
DES J0338$-$2643(Y1,2,3) & 3.3^{+0.8}_{-0.9} & 0.9^{+1.0}_{-1.1} & 5.9^{+1.2}_{-1.1} & 1.54^{+0.39}_{-0.41} & 0 & Y \\
DES J0340$-$2740(Y1) & 3.7^{+1.3}_{-1.2} & 4.3^{+1.1}_{-1.0} & 5.6^{+0.6}_{-1.0} & 1.96^{+0.28}_{-0.52} & 0 & Y \\
DES J0338$-$2614(Y3) & 0.7^{+0.6}_{-0.6} & 1.1^{+0.6}_{-0.7} & 1.1^{+0.7}_{-0.7} & 0.40^{+0.24}_{-0.32} & 0 & N \\
DES J0330$-$2712(SV) & 1.4^{+1.9}_{-1.2} & 2.3^{+1.8}_{-1.2} & 3.2^{+1.8}_{-1.7} & 1.03^{+0.55}_{-0.47} & 0 & Y \\
DES J0328$-$2817(Y3,4) & 2.5^{+0.7}_{-0.7} & 4.8^{+0.9}_{-1.0} & 6.7^{+1.2}_{-1.2} & 2.46^{+0.44}_{-0.48} & 0 & Y \\
DES J0332$-$2847(Y4) & 0.8^{+0.6}_{-0.5} & 1.9^{+0.5}_{-0.6} & 2.2^{+0.7}_{-0.6} & 0.86^{+0.24}_{-0.23} & 0 & N \\
DES J0337$-$2949(Y4) & 2.0^{+1.3}_{-1.2} & 2.5^{+0.9}_{-0.9} & NaN & 1.51^{+0.58}_{-0.58} & 0 & N \\
DES J0332$-$2853(Y4) & 3.1^{+0.9}_{-1.0} & 3.9^{+0.8}_{-0.8} & 3.0^{+0.8}_{-0.8} & 1.40^{+0.32}_{-0.33} & 0 & Y \\
DES J0333$-$2859(Y4) & 0.6^{+0.6}_{-0.4} & 3.3^{+0.6}_{-0.7} & 3.7^{+0.5}_{-0.6} & 1.26^{+0.19}_{-0.37} & 0 & Y \\
DES J0330$-$2749(Y4) & 1.7^{+1.7}_{-1.6} & 2.5^{+1.7}_{-1.6} & NaN & 1.67^{+1.13}_{-1.10} & 0 & N \\
DES J0332$-$2739(Y2,3,4) & 1.4^{+1.3}_{-1.2} & 3.1^{+1.5}_{-1.5} & 3.3^{+1.4}_{-1.3} & 1.28^{+0.47}_{-0.47} & 0 & Y \\
\\ \tableline \\
DES J0627$-$5829(SV) & 1.8^{+1.9}_{-1.1} & 2.0^{+1.9}_{-1.1} & 2.0^{+1.9}_{-1.0} & 0.15^{+0.11}_{-0.12} & 23 & Y \\
DES J0330$-$2732(Y1) & 1.1^{+1.4}_{-1.3} & 1.8^{+0.9}_{-0.8} & 2.9^{+0.7}_{-0.6} & 0.97^{+0.18}_{-0.15} & 2 & N \\
DES J0334$-$2743(Y2,3) & 1.5^{+0.6}_{-0.6} & 4.5^{+0.6}_{-0.7} & 3.9^{+0.7}_{-0.6} & 1.66^{+0.13}_{-0.28} & 12 & N \\
DES J0328$-$2738(Y1,2) & 3.7^{+1.7}_{-1.3} & 7.7^{+1.1}_{-1.2} & NaN & 4.78^{+1.31}_{-0.94} & 2 & Y \\
DES J0059+0006(Y2) & 3.4^{+0.4}_{-2.0} & 2.4^{+0.4}_{-0.9} & 2.7^{+0.6}_{-0.3} & 0.78^{+0.05}_{-0.07} & 2 & N
\enddata
\tablecomments{\scriptsize Columns (2)-(4) give the JAVELIN \textit{r}, \textit{i} and \textit{z} band lags relative to the \textit{g} band and their $1\sigma$ uncertainties. ``NaN'' values mean that we do have good lag measurements in this band. Column (5) gives the accretion disk sizes from the JAVELIN Thin Disk model with $1\sigma$ error bars. Column (6) gives the flags indicating the issues to note for the object. $flag=0$ means no issue to note. $flag=1$ means that the object has lag measurements in multiple seasons that are not consistent with each other. $flag=2$ means we cannot obtain good lags for the object with the ICCF method (see Section \ref{subsec:anl-iccf}). $flag=3$ indicates the lightcurve of the object are not likely to provide good lag measurements given its cadence and depth based on simulation (see Section \ref{subsec:anl-simulation}). Column (7) gives whether the lag signal can be visually seen from the lightcurve of the object, where ``Y'' means the lag is visible while ``N'' means the lag is not clear in the lightcurve (see Section \ref{subsec:anl-lcvisualinspec}).}
\end{deluxetable}

As is shown in Figure \ref{fig:obs_chi2_nexp}, all of the quasars in the main sample have either a large $\chi_r^2$, implicating significant variability relative to the photometric errors, or a large number of epochs, corresponding to a high observational cadence. Most candidates in the SDSS fields have smaller $\chi_r^2$ than those in the main sample, which may explain why only one quasar from the SDSS fields is in the main sample, even though these fields have so many quasars.

\begin{figure*}%[ht!]
%\figurenum{1}
\gridline{\fig{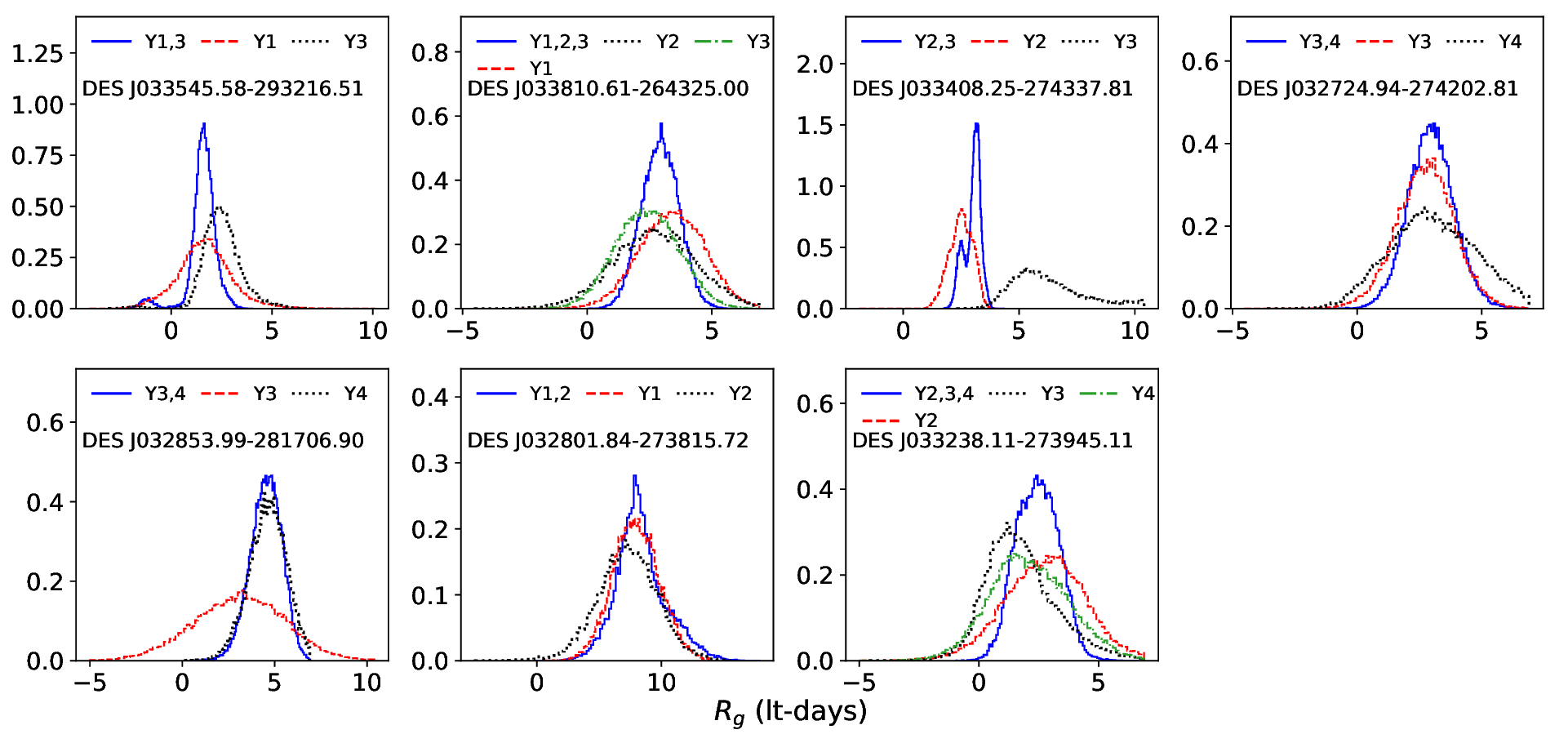}{1.0\textwidth}{}}
%\epsscale{2.0}
\figcaption{Probability distributions for the observed frame \textit{g}-band accretion disk sizes from the JAVELIN Thin Disk model for objects with multi-season lag measurements. In each panel the blue solid line represents the results of a simultaneous fit to all observational seasons with lag measurements, while the red dashed, black dotted and green dash-dotted lines represent the results for individual seasons.\label{fig:alpha_multiseason}}
\end{figure*}

We treat the observations in each season as independent time series and analyze them separately with JAVELIN. Column (1) in Table \ref{tab:objresult} shows the seasons where we obtain good measurements for each object. There are seven objects that have good measurements in multiple seasons. We compare the accretion disk sizes from different seasons in Figure \ref{fig:alpha_multiseason}. Most objects show consistent disk sizes from different seasons at the 1$\sigma$ level, while only one object (DES J033408.25$-$274337.81) shows different disk sizes from the different seasons. For these objects, we show the lags and disk sizes from fitting the lightcurves in all seasons simultaneously in Table \ref{tab:objresult} and adopt them for further analysis. We add $flag=1$ to DES J033408.25$-$274337.81 in Table \ref{tab:objresult}. The discrepancy in the disk sizes from different seasons may be because the accretion disk undergoes structural changes between seasons, or that the time lags between different photometric bands are not exactly described by the simple scenario that we assumed. 

A successful lag measurement in one season does not guarantee successful lag measurements in other seasons. There are three factors that can prevent a lag measurement in a given season: the AGN variability, the lightcurve photometric data quality and the observational cadence. A successful lag measurement, and particularly one easily confirmed by visual inspection, requires a significant flux variation relative to the photometric errors that is also well-sampled by the observational cadence. AGN variability is sufficiently stochastic that the largest photometric variations may fall outside of the range of the lightcurve data, be on order the photometric noise (or less), or be insufficiently well sampled to identify a clear signal. For example, less than a quarter of the objects in the MaxVis and SDSS fields meet our selection criteria (Section \ref{sec:analysis}) in more than one season, while the other seasons do not have enough epochs and/or sufficient lightcurve data quality (signal-to-noise ratio).

\begin{figure}%[ht!]
%\figurenum{1}
\plotone{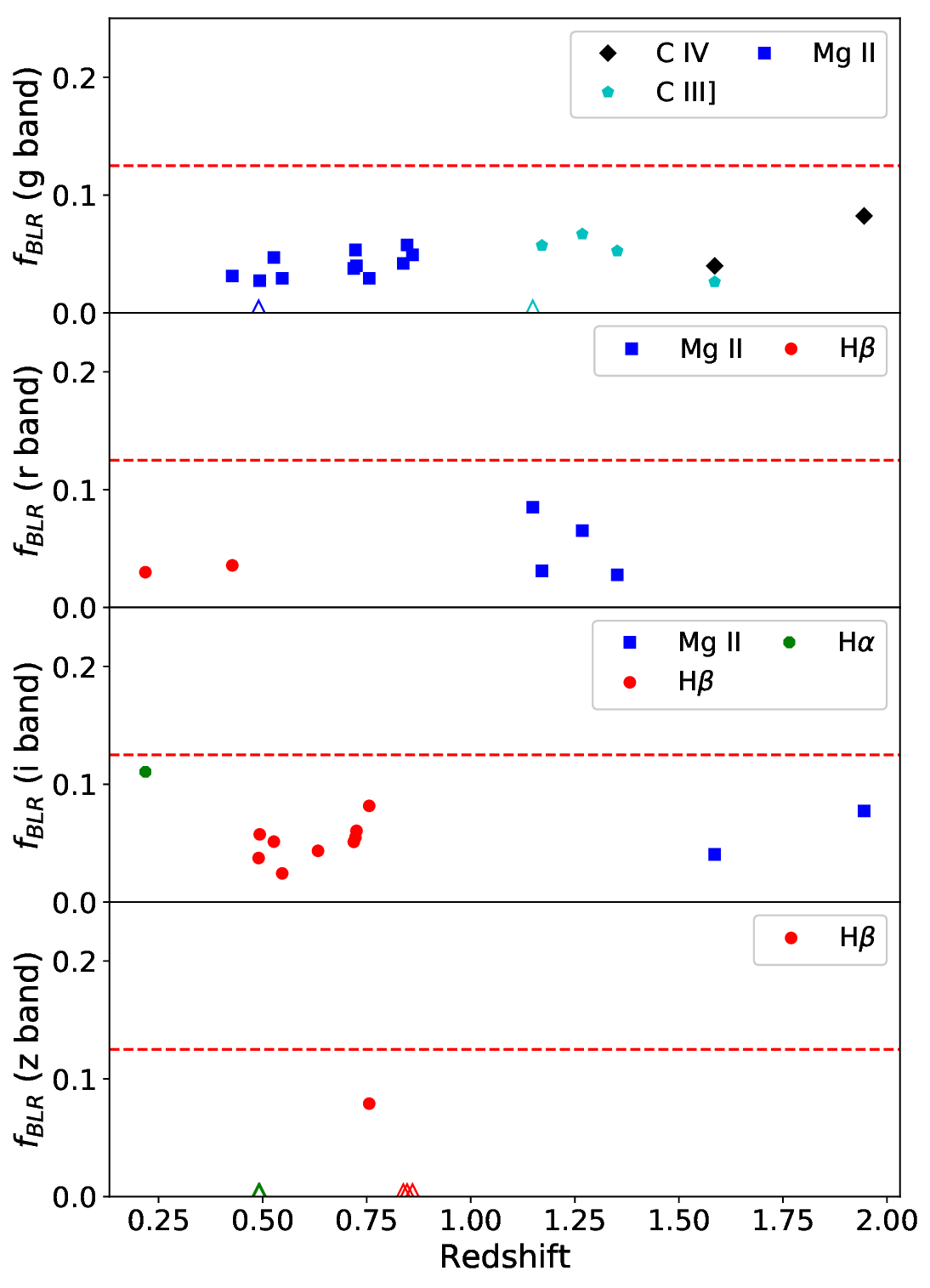}
\figcaption{Broad line contamination fraction $f_{BLR}$ as a function of redshift. $f_{BLR}$ is the ratio of the equivalent width of the emission line to the effective width of the broad band filter. The panels from top to bottom represent the broad line contamination in the \textit{g}, \textit{r}, \textit{i} and \textit{z} bands, respectively. The filled symbols represent the emission lines that could contaminate the continuum lag measurements for the object, with different colors and shapes for different emission lines. The empty triangles are the emission lines where we did not derive $f_{BLR}$ because the line falls out of the wavelength range of the spectrum or falls into a region where the spectrum is too noisy. The red dashed line is drawn at $f_{BLR} = 12.5\%$.\label{fig:contam_bel}}
\end{figure}

The variability of the broad emission lines can contaminate the continuum lag measurements. The variations of broad emission lines lag the continuum due to the light travel time from the accretion disk to the BLR. The broad line variability may consequently make the measured lag larger than the real continuum lag. We assess the contamination from broad emission lines as the ratio of the equivalent width of the emission line to the effective width of the broad band filter, referred to as $f_{BLR}$, similar to \citet{Homayouni2018}. Those authors identified potential contamination if $f_{BLR}>12.5\%$. For each object in the main sample in the MaxVis and the C26202 fields, we calculated the $f_{BLR}$ for the Ly$\alpha$, C IV, C III], Mg II, H$\beta$ and H$\alpha$ if the line fell into the band pass of any of the \textit{g}, \textit{r}, \textit{i} or \textit{z} bands. For the object within the SDSS footprint, we adopted the equivalent width of the emission lines from \citet{Shen2011}. We show $f_{BLR}$ as a function of redshift for the \textit{g}, \textit{r}, \textit{i} and \textit{z} bands in Figure \ref{fig:contam_bel}. All objects show $f_{BLR}$ less than 12.5\%, indicating that our continuum lag measurements have little contamination from broad lines. There are a few lines where we did not derive $f_{BLR}$ because the line falls out of the wavelength range of the OzDES spectrum from 3700 \AA\, to 8800 \AA\, or falls into a region where the spectrum is too noisy. These cases are indicated by the empty triangles in Figure \ref{fig:contam_bel}. However, it is clear from Figure \ref{fig:contam_bel} that no emission line has sufficiently large flux that would contaminate the lag measurements, so it is unlikely that the few quasars with unmeasured $f_{BLR}$ would significantly affect our conclusions. 

%Analysis-ICCF
\subsection{ICCF Analysis} \label{subsec:anl-iccf}
We also use the conventional ICCF method \citep[e.g.][]{Peterson1998,Peterson2004} to derive time lags with the public code PyCCF \citep{pyccf}. It cross-correlates two lightcurves using linear interpolation and measures the peak location $\tau_{peak}$ and centroid location $\tau_{cent}$ of the cross-correlation function (CCF) using points with cross-correlation coefficients $r\geq 0.8r_{peak}$, where $r_{peak}$ is the peak value of the CCF. To estimate the uncertainty of the lag measurements, PyCCF creates a series of independent realizations of the lightcurve through Monte Carlo iterations with flux randomization and random subset selection (with replacement, also known as ``bootstrapping''), and builds up the cross-correlation centroid distribution (CCCD) and cross-correlation peak distribution (CCPD). We create 20000 realizations of the lightcurve and set the threshold of a ``significant'' correlation to be 0.5, i.e. realizations with $r_{peak} \le 0.5$ are excluded from CCCD and CCPD. The median fraction of the failed realizations is 14\% for the main sample, except 5 objects that show failure fraction larger than 70\% in at least two bands. The objects with large failure fractions are all flagged objects, or objects with lags significantly smaller than the observational cadence where ICCF is known to have trouble recovering lags \citep[e.g.][]{Jiang2017}. 

Figures \ref{fig:lagclean1} - \ref{fig:lagflag1} show the CCCD and CCPD compared to the JAVELIN lag distributions. For objects in Figures \ref{fig:lagclean1} - \ref{fig:lagclean3}, the ICCF results are generally consistent with the JAVELIN results in at least one of the \textit{r}, \textit{i} and \textit{z} bands, while the lag distributions from ICCF are significantly wider than the JAVELIN lag distributions. However, we also find a few objects where ICCF does not produce good lag measurements, or the ICCF results deviate significantly from the JAVELIN results in all bands. We add $flag=2$ to these objects in Table \ref{tab:objresult}, and show their JAVELIN and ICCF results in Figure \ref{fig:lagflag1}. Finally, we note that we reanalyzed DES J033719.99$-$262418.83 from \citet{Mudd17}, which is in the C2 supernova field. This object is a marginal detection both here and in \citet{Mudd17}, with large uncertainties in lags from JAVELIN and ICCF, and we therefore do not include it in our main sample. We provide more detailed comments on the comparison between the JAVELIN and ICCF results in the Appendix.

The larger uncertainties of lag measurements from ICCF compared to JAVELIN are typical. JAVELIN likely underestimates uncertainties because of non-Gaussian or other issues in the lightcurve uncertainties and because the second lightcurve may not simply be a shifted, scaled and smoothed version of the first. On the other hand, previous studies \citep[e.g.][]{Jiang2017} find that ICCF does not work well in recovering time lags less than the cadence of the lightcurve, which is the case of many of our objects in the C26202 field. JAVELIN provides a much better means of interpolating and weighting interpolated points than linear interpolation. We further compare the performance of JAVELIN and ICCF through simulations in Section \ref{subsec:anl-simulation}. We only report the lags from JAVELIN hereafter.

%Analysis-Visual Inspection of Lightcurves
\subsection{Visual Inspection of Lightcurves} \label{subsec:anl-lcvisualinspec}
We visually inspected the lightcurves of all objects in the main sample to see whether we can identify the lag signals by eye. Figure \ref{fig:lcexamp} shows examples of the visible lag signals marked by the arrows. The local peaks or valleys pointed to by the arrows visibly appear later in the bands with longer wavelengths. We also marked the visible lag signals in the full-size lightcurve plots presented in the online journal. 

We find that 12 out of the 17 unflagged objects and 2 out of the 5 flagged objects in the main sample show at least one visible lag feature in at least one of the \textit{griz} bands. These results are listed in Table \ref{tab:objresult}. We note that most of the unflagged objects in the main sample have lags that are directly visible from the lightcurves. The objects where the lag signals are not clearly visible often have lags that are smaller than the cadence of the lightcurve. This is true of a larger fraction of the flagged sample.
%\begin{figure}%[ht!]
%\figurenum{1}
%\plotone{LC/lczoom_MaxVis-113_C26202-33.eps}
%\figcaption{Zoom-in lightcurves of DES J063037.48$-$575610.30 and DES J034003.89$-$264524.52. The symbols have the same meaning as those in Figure %\ref{fig:lcgood1}. The black rectangle represents the region where the lag signal is most apparent. The red arrows point to the approximate positions of %the features that show visible lags.\label{fig:lczoom}}
%\end{figure}

%Verification of Lag Measurements 
\section{Verification of Lag Measurements} \label{sec:verification}

%Analysis-Simulation
\subsection{Simulations} \label{subsec:anl-simulation}

We ran simulations to further verify the lag measurements. First we created simulated lightcurves with just Gaussian noise. For each object in the main sample, we calculated its mean magnitude $\mu_m$ and standard deviation $\sigma_m$ within an observational season in each band. For each season and band, we created a simulated lightcurve where each epoch is a Gaussian deviation of dispersion $\sigma_m$ about $\mu_m$ with the same sampling as the observed lightcurve. We set the uncertainty of each simulated data point to be the same as the uncertainty of the corresponding data point in the observed lightcurve. In this case, we created simulated lightcurves that have the same cadence as the observed lightcurve, but do not have lags. We then look for the lags between the \textit{observed} \textit{g}-band lightcurve and the \textit{simulated} \textit{r}, \textit{i}, \textit{z} band lightcurves. We find that the probability distributions of the time lags from JAVELIN have multiple peaks of both signs with no evidence of a clear, positive lag. This indicates that JAVELIN does not produce fake detections from lightcurves with no lag signal. 

We then created simulated lightcurves from the DRW model. For each object in the main sample, we constructed DRW lightcurves with a 0.05-day cadence using the best-fit $\sigma_{\mbox{\tiny DRW}}$ and $\tau_{\mbox{\tiny DRW}}$ from JAVELIN for the \textit{g} band. We shifted the \textit{g}-band lightcurve by a time lag of 1, 2, 4 or 8 days to create simulated lightcurves in the \textit{r}, \textit{i} and \textit{z} bands. We did not add further noise to the shifted lightcurves. We created five realizations of the DRW lightcurve for each input time lag. We sampled the 0.05-day cadence simulated lightcurves to the same cadence as the observed lightcurves, and set the photometric uncertainty of each data point to be the same as the corresponding data point in the observed lightcurve. 

We ran JAVELIN on the simulated lightcurves to check whether JAVELIN can reproduce the input time lag. The first row of Figure \ref{fig:sim_allcomp} shows an example of the JAVELIN results from fitting the simulated DRW lightcurves of DES J063037.48$-$575610.30, a quasar in the MaxVis field with a $\sim$ 1-day observational cadence. Figure \ref{fig:sim_allcomp} shows that JAVELIN can reproduce the input time lags at the 1$\sigma$ level in most realizations, although there are a few cases where the JAVELIN lags deviate significantly from the input. We performed these simulations for all 22 quasars in the main sample. For only one object we did not reproduce the input lag in most realizations. We add $flag=3$ to this object in Table \ref{tab:objresult}. 

In addition to JAVELIN, we also test the ICCF method on the simulated DRW lightcurves. Figure \ref{fig:sim_drw_iccf} shows the lag distributions for the simulated DRW lightcurves of DES J063037.48$-$575610.30 (same quasar as Figure \ref{fig:sim_allcomp}) with the ICCF method. It shows that the ICCF lags are also usually consistent with the input, although the uncertainty of the ICCF lags, and the number of cases where the ICCF lag deviates significantly from the input, is larger than for the JAVELIN lags. For most quasars in the main sample, the simulation results for the ICCF method are similar to Figure \ref{fig:sim_drw_iccf}. However, we note a few cases where the ICCF method cannot reproduce the input lags at all. For instance, the lower panel of Figure \ref{fig:sim_drw_iccfbad} shows the ICCF lags from the simulated lightcurves of DES J032801.84$-$273815.72 with a 8-day input lag. The lag distributions are wide with multiple peaks, or show a peak that deviates significantly from the input lag, indicating that the ICCF method is unlikely to provide reliable lag measurements for this quasar, while the upper panel of Figure \ref{fig:sim_drw_iccfbad} shows that JAVELIN can recover the input time lag from the simulated lightcurves of this quasar. Note that in Section \ref{subsec:anl-iccf} we flag DES J032801.84$-$273815.72 because it does not have good ICCF lags from the \textit{actual} lightcurves. Based on these simulation results, we conclude that objects with $flag=2$ in Table \ref{tab:objresult} are not necessarily unreliable objects. It may be that ICCF does not work as well as JAVELIN for some values of the cadence and variability. 

\subsubsection{Deviations from Pure DRW} \label{ssubsec:sim_nondrw}
The observed lightcurves can deviate from the DRW model due to several factors. First, the observed lightcurves are not noiseless. To account for this effect, we added noise to our simulated DRW lightcurves so that each epoch is a Gaussian deviation of dispersion $\sigma_m$ about the original noiseless value, where $\sigma_m$ is the uncertainty of the simulated epoch and is the same as the uncertainty of the corresponding observed epoch. We ran JAVELIN on these noisy simulated lightcurves and the second row of Figure \ref{fig:sim_allcomp} shows an example of the results. While the noise leads to a few secondary peaks, the lag distributions of most realizations still concentrate around the input lag. This indicates that we are still able to recover the lags with JAVELIN with the signal-to-noise level of the observed lightcurves. We also created simulated lightcurves with two times and four times the observed noise. In these cases, JAVELIN produces flatter lag distributions with larger noise, and is generally not able to recover the input lag once the noise is as large as four times the observed value.

Some studies of the AGN variability found a steeper power spectral density (PSD) than the DRW model at short time scales using Kepler data \citep[e.g.][]{Mushotzky2011,Smith2018}. We define a ``Kepler-exponential'' (KE) covariance function to include this effect in the simulation:
\begin{equation}
S(\Delta t) = \sigma^2 \, [(1 + C) \, {\rm exp}(-|\Delta t / t_1|) - C \, {\rm exp}(-|\Delta t / t_2|)]
\label{eq:cov_ke}
\end{equation}
where $C = t_2 / (t_1 - t_2)$. The $\sigma$ and $t_1$ are equivalent to the $\sigma_{\mbox{\tiny DRW}}$ and $\tau_{\mbox{\tiny DRW}}$ in the DRW model, while we can vary $t_2$ to make a cut-off at short time scales as found in the Kepler lightcurves. We adjusted $t_2$ so that the structure function and the PSD start to deviate from DRW at around 30 days and created simulated lightcurves with the KE covariance function. The third row of Figure \ref{fig:sim_allcomp} shows an example of the JAVELIN results from fitting the KE lightcurves. We note that JAVELIN can still recover the input time lags. This indicates that the deviations from DRW at short times have little impact on our results and conclusions. 

\subsubsection{Diffuse Continuum from BLR} \label{ssubsec:sim_dc}
\begin{figure*}%[ht!]
\gridline{\fig{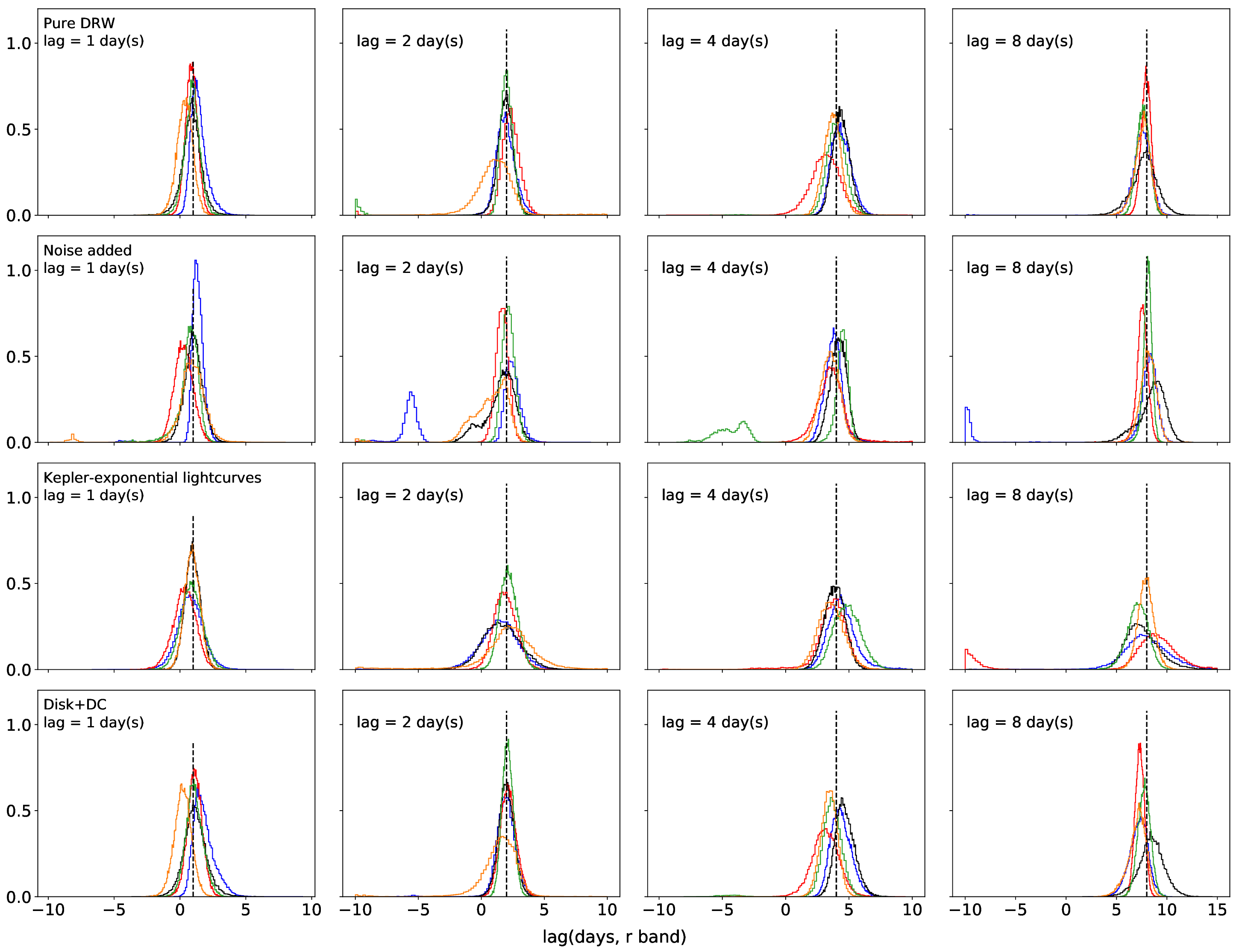}{1.0\textwidth}{}}
%\epsscale{2.0}
\figcaption{Probability distribution of \textit{r}-band time lags relative to \textit{g}-band from fitting the five realizations (LC0 - LC4) of the simulated light curves of DES J063037.48$-$575610.30 in the MaxVis field. The histograms with different colors in each panel represent the JAVELIN result for different realizations of the simulated lightcurves. The black dashed line in each panel represents the position of the input time lag, which is 1, 2, 4 and 8 day(s) for the first, second, third and fourth column, respectively. The first through the fourth row represent the results from pure DRW lightcurves, DRW lightcurves with noise, Kepler-exponential lightcurves and the composite lightcurves of the accretion disk and the diffuse continuum from the BLR.\label{fig:sim_allcomp}}
\end{figure*}

\begin{figure*}%[ht!]
\gridline{\fig{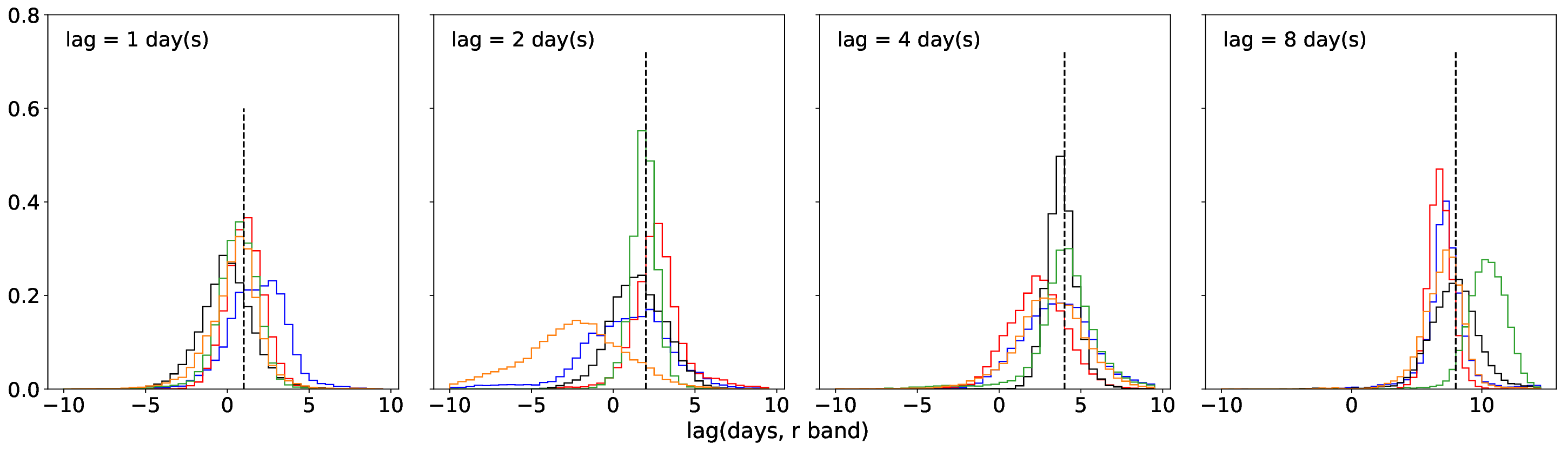}{1.0\textwidth}{}}
%\epsscale{2.0}
\figcaption{Probability distribution of \textit{r}-band time lags relative to \textit{g}-band from fitting the five realizations (LC0 - LC4) of the simulated DRW light curves with the same $\sim$ 1 day cadence as DES J063037.48$-$575610.30 (same quasar as Figure \ref{fig:sim_allcomp}) with ICCF. The histograms with different colors in each panel represent the ICCF center distributions for different realizations of the DRW lightcurves. The black dashed line in each panel represents the position of the input time lag, which is 1, 2, 4 and 8 day(s) for the first, second, third and fourth column, respectively.\label{fig:sim_drw_iccf}}
\end{figure*}

\begin{figure}%[ht!]
%\figurenum{1}
\plotone{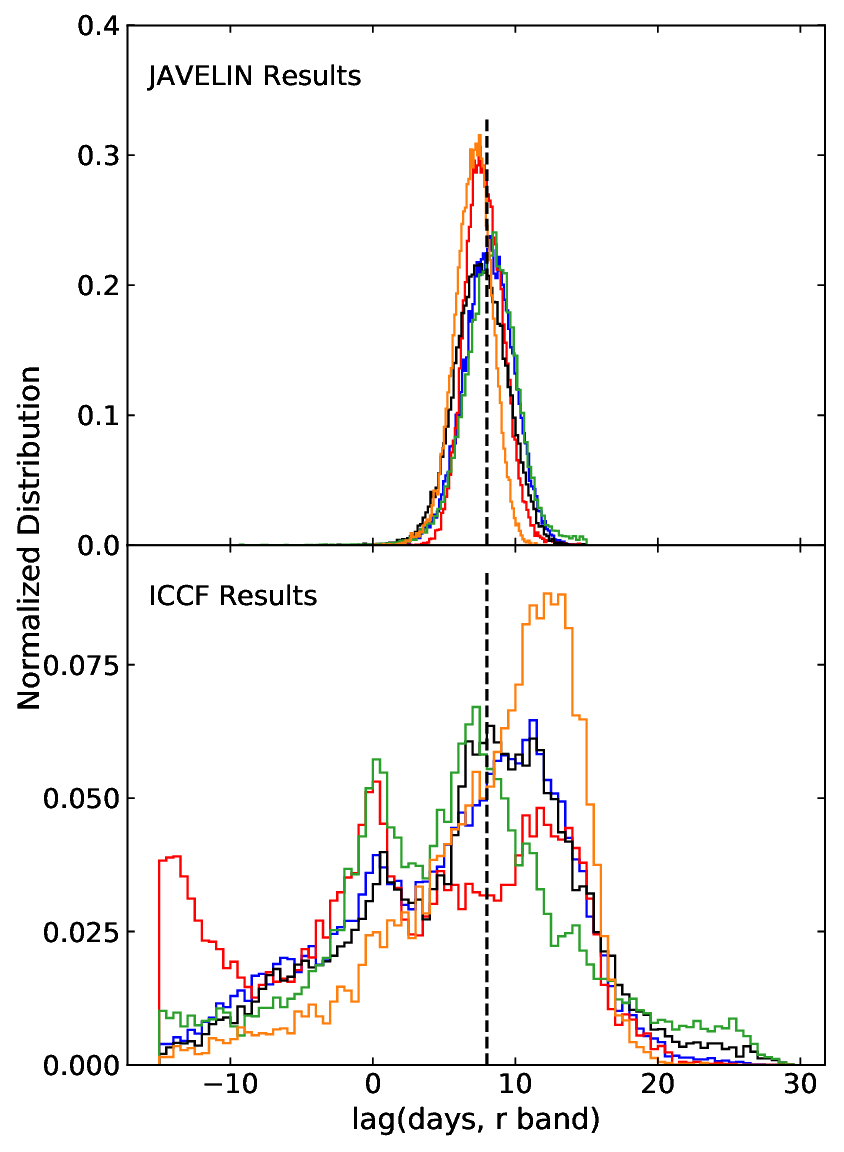}
\figcaption{Probability distribution of \textit{r}-band time lags relative to \textit{g}-band from fitting the simulated DRW lightcurves of DES J032801.84$-$273815.72 with JAVELIN (upper panel) and ICCF(lower panel). The histograms with different colors in each panel represent the JAVELIN or ICCF results for different realizations of the DRW lightcurves. The black dashed line in each panel represents the position of the 8-day input time lag.\label{fig:sim_drw_iccfbad}}
\end{figure}

The diffuse continuum from the BLR is a potential contamination source for the disk continuum lightcurves and may lead to an overestimation of the disk size \citep[e.g.][]{Korista2001,Lawther2018,Cackett2018}. We assess the possible contamination from the diffuse continuum by composite simulated lightcurves of the accretion disk and the diffuse continuum. We first estimate the fraction of the diffuse continuum in the total continuum using ``Model 1'' in Figure 8 of \citet{Lawther2018}. We then calculate the emission line lags of our objects using the $R_{BLR}$ - Luminosity relation and the flux-calibrated OzDES spectra, and estimate the diffuse continuum lags using the scaling between the line lags and diffuse continuum lags from ``Model 1'' in Table 2 and Figure 8 of \citet{Lawther2018}. We shift the driving continuum lightcurve created in Section \ref{subsec:anl-simulation} by the diffuse continuum lags to create the diffuse continuum lightcurves and sampled these lightcurves to the observed epochs. We create the composite lightcurves by 
\begin{equation}
F_{comp,i} = (f_{dc}\times CF)F_{dc,i} + (1 - f_{dc}\times CF)F_{disk,i}
\end{equation}
where $F_{comp,i}$ is the flux of the \textit{i}th epoch of the composite lightcurves, $f_{dc}$ is the diffuse continuum fraction, $CF$ is the covering factor of the BLR and $F_{dc,i}$ and $F_{disk,i}$ are the flux of the \textit{i}th epoch of the diffuse continuum lightcurves and the accretion disk lightcurves, respectively. Here we adopt $CF=0.4$ \citep[e.g.][]{Dunn2007,Lawther2018}. The bottom row of Figure \ref{fig:sim_allcomp} shows an example of JAVELIN results from the composite lightcurves. We do not find any significant systematic shift of the lags relative to the lags from the pure disk lightcurves after adding the diffuse continuum component. This indicates that our disk lag measurements have little contamination from the diffuse continuum.

%Verification - Re-weighting
\subsection{Re-weighting the Lightcurves} \label{subsec:veri-rew}
\begin{figure*}%[ht!]
\gridline{\fig{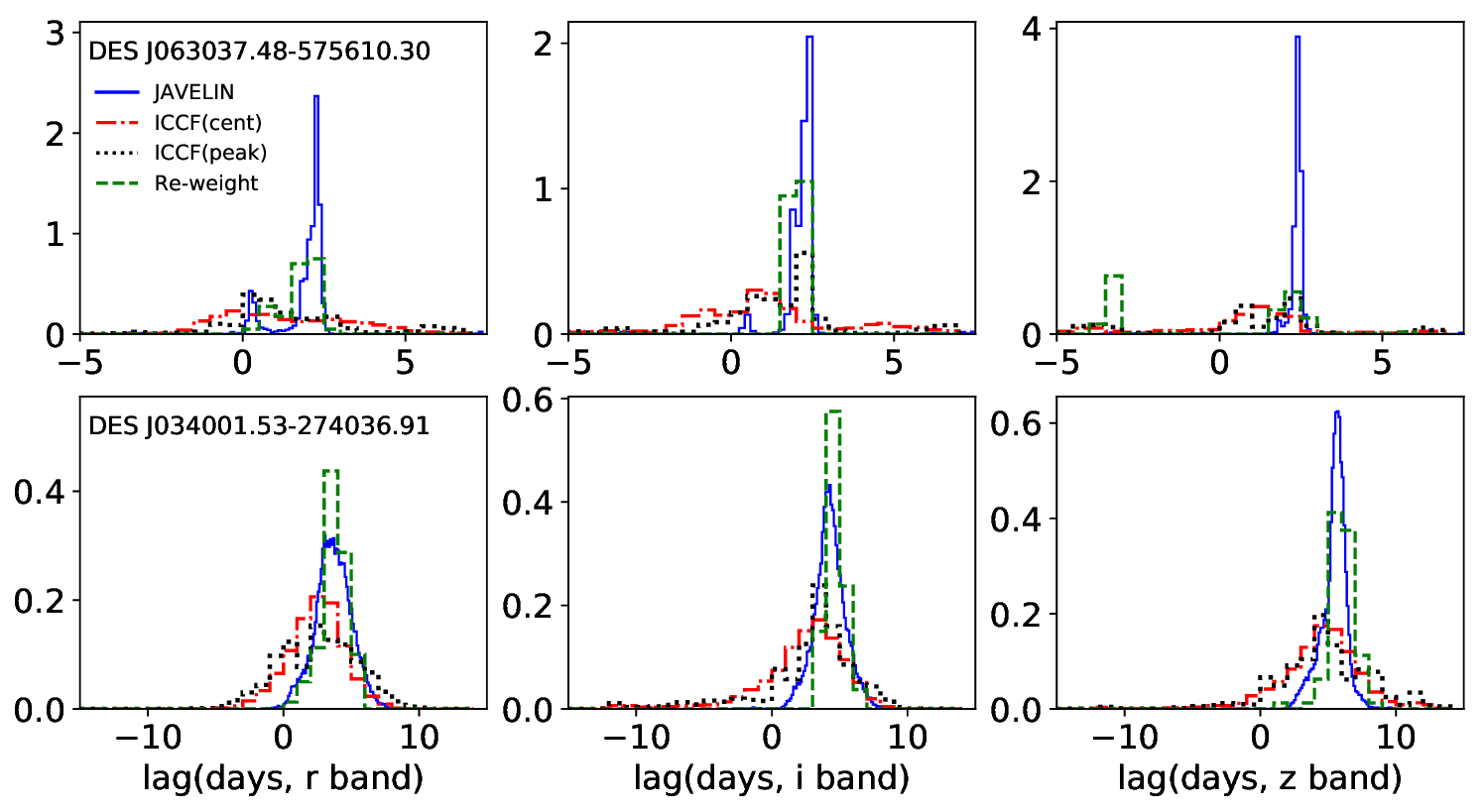}{1.0\textwidth}{}}
%\epsscale{2.0}
\figcaption{Probability distribution of time lags in the \textit{r}, \textit{i}, \textit{z} band relative to \textit{g} band for DES J063037.48$-$575610.30 (upper row) and DES J034001.53$-$274036.91(lower row). The blue solid, red dash-dotted and black dotted lines represent the results from JAVELIN, the ICCF center distribution and the ICCF peak distribution, respectively. The green dashed line represents the distribution of the median JAVELIN lags from fitting the re-weighted lightcurves. \label{fig:veri-rew}}
\end{figure*}

To further verify the values and uncertainties of the lags from JAVELIN, we adopt a ``bootstrap'' like method. For a lightcurve in band X (X=\textit{g},\textit{r},\textit{i},\textit{z}) with $N_{X}$ data points, we randomly pick $N_{X}$ points with replacements. If a data point is picked $N_{pick}$ times, we divide its errorbar by $\sqrt{N_{pick}}$. If a data point is not picked, we double its errorbar. We do not simply exclude the data point like the traditional ``bootstrap'' method, since the cadence is critical to the time series analysis, while in the traditional ``bootstrap'' there is no equivalent to the ``time'' axis for the lag measurements. Doubling the errorbar significantly reduces the weight in the data point, which does similar job as removing the data point without qualitatively changing the lightcurve. We created 80 re-weighted lightcurves for a few representative objects in the main sample, and we ran JAVELIN on each of the re-weighted lightcurves. We obtain the median JAVELIN lag of each re-weighted lightcurve, and compare its distribution to the previous lag distributions from JAVELIN and ICCF. Figure \ref{fig:veri-rew} shows the results of DES J063037.48$-$575610.30 from the MaxVis field in the upper row and DES J034001.53$-$274036.91 from the C26202 fields in the lower row. The median lag distributions from the re-weighted lightcurves are generally consistent with the previous JAVELIN and ICCF results. The \textit{z}-band lag distribution of DES J063037.48$-$575610.30 from the re-weighted lightcurves show a small secondary peak near $-5$ days, possibly due to the common $\sim$ 5-day gaps in its lightcurve. The positive lag distributions are still consistent with previous lag distributions. These results again verify our lag measurements in Section \ref{sec:analysis}.

%Verification - Gaussianity 
\subsection{``Gaussianity'' of the Photometric Errors} \label{subsec:veri-gauss}
One assumption of JAVELIN is that the input errors are Gaussian. We therefore assess the ``Gaussianity'' of the photometric errors from DES with the standard stars in the SDSS Stripe 82 fields. We select a subsample of the standard stars from the SDSS Stripe 82 standard star catalog by \citet{Ivezic2007} with the same distribution of \textit{g}-band magnitudes as the whole standard star sample. We construct DES lightcurves for the stars following the same process as our quasar sample, and exclude the stars that have less than 200 epochs over the five DES observational seasons or have unreliable photometries based on DES flags. For each standard star, we calculate the ratio $(m_{i,X} - \overline{m}_X)/\sigma_{i,X}$ for each data point in the lightcurve, where $m_{i,X}$ and $\sigma_{i,X}$ represents the magnitude and magnitude error of the \textit{i}th data point in band X (X=\textit{g},\textit{r},\textit{i},\textit{z},\textit{Y}), and $\overline{m}_X$ represents the mean magnitude in band X. Assuming that the standard stars from \citet{Ivezic2007} are non-variable objects, the distribution of the ratio $(m_{i,X} - \overline{m}_X)/\sigma_{i,X}$ should follow a Gaussian distribution centered at 0 with a standard deviation equal to 1. In addition, we calculate $\chi_r^2$ defined in Section \ref{sec:analysis} for the whole SV - Y3 period for each standard star. If the photometric errors are well estimated, $\chi_r^2$ should be close to 1. We do not include the Y4 data in this section, since the calibrations for the Y4 data in the MaxVis field and the SDSS fields differ from the DES FGCM calibration \citep{Burke2018} for other seasons and fields, and none of the lag measurements in the main sample are from the Y4 data in these fields. 
 
\begin{figure}%[ht!]
%\figurenum{1}
\plotone{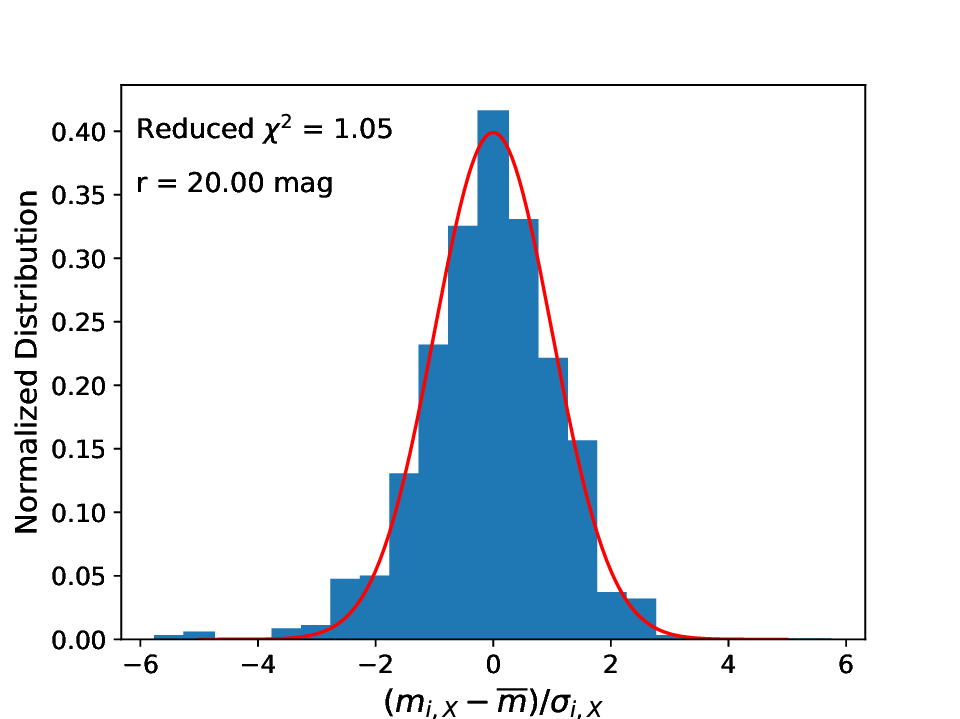}
\figcaption{Example of the ``Gaussianity'' of the DES photometric errors. The blue histograms show the distribution of $(m_{i,X} - \overline{m}_X)/\sigma_{i,X}$ for a standard star. The red line shows a Gaussian profile centered at 0 with a standard deviation equal to 1. The upper left corner shows $\chi_r^2$ value (see Section \ref{subsec:veri-gauss}) and the \textit{r}-band magnitude of the star.\label{fig:veri_gauss_examp}}
\end{figure}

\begin{figure}%[ht!]
%\figurenum{1}
\plotone{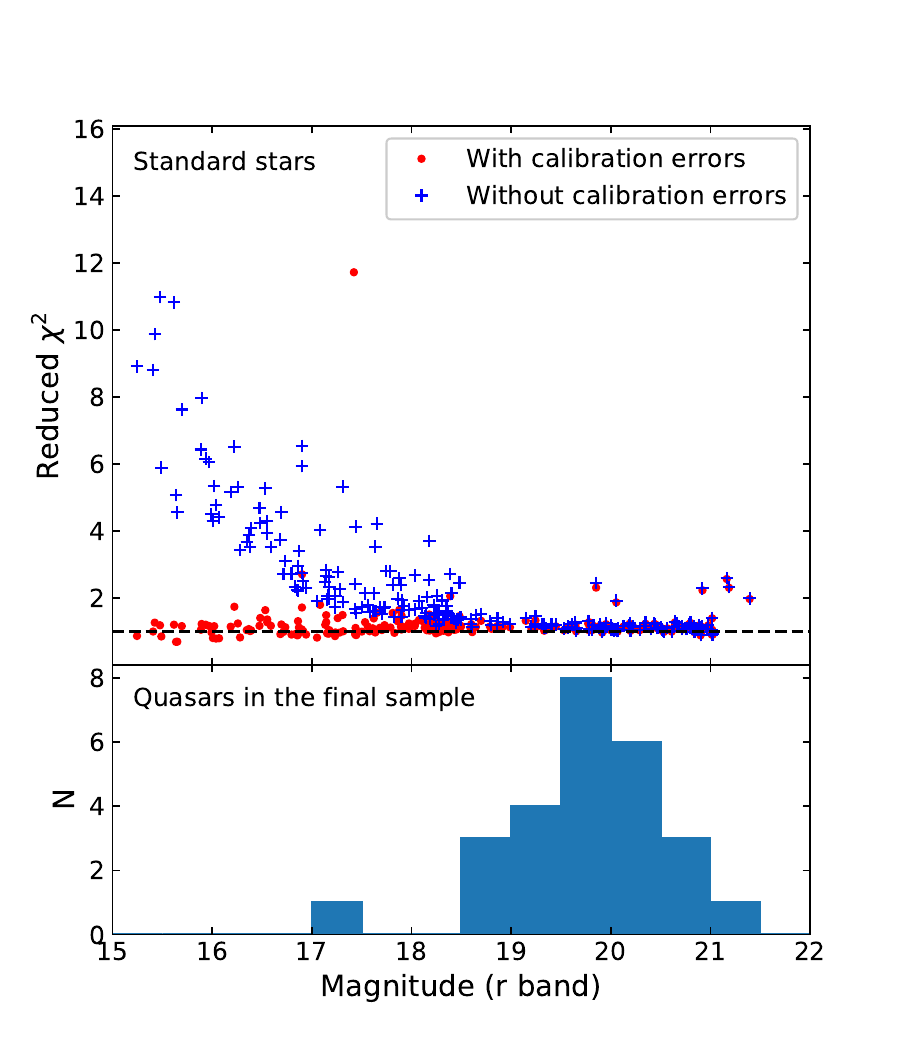}
\figcaption{(upper panel) $\chi_r^2$ (see Section \ref{subsec:veri-gauss}) as a function of the \textit{r}-band magnitude for the standard stars. The blue crosses show $\chi_r^2$ where we do not consider the DES calibration errors, while the red points represent the cases where we add the DES calibration errors following the same process as the quasar lightcurves. The black dashed line represents the position where $\chi_r^2$ equals 1. (lower panel) Magnitude distribution of the quasars in the main sample.\label{fig:veri_gauss_chi2}}
\end{figure}

Figure \ref{fig:veri_gauss_examp} shows an example of the distribution of $(m_{i,X} - \overline{m}_X)/\sigma_{i,X}$ for a standard star with $\chi_r^2$ around 1.05. The distribution agrees well with the superimposed Gaussian profile, indicating that the DES photometric errors are consistent with Gaussian errors in this case. Figure \ref{fig:veri_gauss_chi2} shows $\chi_r^2$ of the standard stars as a function of the \textit{r}-band magnitude in the upper panel, and the distribution of the \textit{r}-band magnitude of the main quasar sample in the lower panel. Most stars within the magnitude range of most quasars in the main sample show $\chi_r^2$ close to 1, indicating that the DES photometric errors for quasars within this magnitude range are well estimated. We note that $\chi_r^2$ is also a good indicator of ``Gaussianity'', and most stars with $\chi_r^2$ close to 1 show similar distributions as Figure \ref{fig:veri_gauss_examp}, so we expect that the DES photometric errors are also close to Gaussian within the magnitude range of the main quasar sample. For bright stars, $\chi_r^2$ becomes significantly larger than 1 if we do not consider the calibration errors, but stays near 1 when we take the calibration errors into account. This indicates that the total photometric uncertainties of these bright stars are dominated by calibration errors. Since the DES standard star observations generally follow the same strategy among the fields we study, we expect the results from the SDSS Stripe 82 fields are applicable to our other fields. We therefore conclude that the photometric errors of the FGCM calibrated DES data are Gaussian and of the correct amplitude.

%Disk size vs. BH mass
\section{Disk Size - Black Hole Mass relation} \label{sec:bhmass_disksize}

For objects within the MaxVis and the C26202 fields, we estimate the SMBH mass of each quasar in the main sample through the single-epoch method using the broad H$\beta$, Mg II or C IV line in the OzDES spectra. We calculate the black hole mass as
\begin{equation}
M_{BH} = f\frac{R_{BLR} \Delta V^2}{G}    
\end{equation}
where $R_{BLR}$ is the size of BLR, $\Delta V$ is the line width of the broad emission line used for the estimation, and $f$ is the dimensionless ``virial factor'' that accounts for other unknown factors, such as the inclination, structure and kinematics of the BLR. 

We use the line dispersion of the broad emission lines as our indicator of the line width, as previous studies have shown this provides more robust black hole mass estimates compared to the Full-Width Half Maximum (FWHM) \citep[e.g.][]{Peterson2004}. The line dispersion is defined as
\begin{equation}
\sigma_{line}^2(\lambda) = \left[\int \lambda^2 P(\lambda) d\lambda \, \Big/ \int P(\lambda) d\lambda \right] - P_0(\lambda)^2
\end{equation}
where $P(\lambda)$ is the line profile and $P_0(\lambda)$ is the first moment of the line profile. We use the public code PySpecKit \citep{pyspeckit} in the analysis of the spectra. For the Mg II line, we also fit and subtract the iron emission lines using the template from \citet{Vestergaard2001}. 

We measure the monochromatic luminosity using the flux-calibrated OzDES spectra \citep{Hoormann2018}, and calculate $R_{BLR}$ using the $R_{BLR}$ - Luminosity relations from \citet{Bentz2013} for H$\beta$, from \citet{McLure2002} for Mg II and from \citet{Hoormann2018} for C IV. We adopt a virial factor of $f=4.31$ from \citet{Grier2013}. For objects within the SDSS fields, we adopt the single-epoch black hole masses from \citet{Shen2011}. The error in single epoch black hole mass is roughly 0.4 dex, based on the uncertainties in the virial factor and the $R_{BLR}$ - Luminosity relation \citep[e.g.][]{Peterson2014}. We do not further consider the uncertainties from the line width and continuum luminosity measurements for individual objects, which are very small compared to 0.4 dex.

In comparison to the observed accretion disk sizes, the disk sizes predicted by the standard thin disk model at effective wavelength $\lambda$ has the analytical form
\begin{equation}
R_{\lambda} = \left[\left(\frac{GM}{8\pi \sigma}\right)\left(\frac{L_{Edd}}{\eta c^2}\right)(3+\kappa)\dot{m}_E \right ]^{1/3} \left( \frac{k\lambda}{hc} \right)^{4/3}
\label{eq:disksize}    
\end{equation}
where $M$ is the mass of the SMBH, $L_{Edd}$ is the Eddington luminosity, $\dot{m}_E$ is the Eddington ratio defined as the bolometric luminosity $L_{Bol}$ divided by the Eddington luminosity, $\eta$ is the radiative efficiency defined as $L_{Bol} = \eta \dot{M} c^2$, where $\dot{M}$ is the mass accretion rate, and $\kappa$ is the ratio of external to internal heating. Note that the disk size in Equation (\ref{eq:disksize}) is defined as the position where $kT(R_{\lambda}) = hc/\lambda$. This in fact assumes that the emission at wavelength $\lambda$ is solely contributed from radius $R$. However, in reality the emission at $\lambda$ also has contributions from other radii. The disk size from continuum reverberation mapping is in fact a flux-weighted mean radius $\langle R_\lambda \rangle$ defined as
\begin{equation}
\langle R_\lambda \rangle = \frac{\int_{R_{min}}^{\infty} B(T(R)) R^2 \, dR}{\int_{R_{min}}^{\infty} B(T(R)) R \, dR} = X R_{
\lambda}
\label{eq:fluxweightedR}
\end{equation}
where $R_{min}$ is the inner edge of the disk and $B(T)$ is the Planck function. Assuming the inner edge $R_{min} \rightarrow 0$, the conversion factor $X=2.49^{4/3}=3.36$ if the disk emission does not vary. \citep[e.g.][]{Fausnaugh2016,Tie2018}.

\citet{Tie2018} found that the conversion factor $X$ will be larger when taking the variation of the disk emission into account. Assuming the ``lamppost'' model \citep[e.g.][]{Cackett2007} for the variability, the temperature fluctuation of the disk is 
\begin{equation}
T(R,t) = T_0(R) [1+f(t-R/c)]     
\end{equation}
where $T_0(R)$ is the unperturbed temperature at radius $R$, and $f(t-R/c)$ is the fractional change of the temperature lagging the variation at the disk center by the light travel time $R/c$. Assuming the temperature fluctuation is small, the fluctuation of the disk surface brightness is 
\begin{equation}
\delta I(R,t) \propto f(t-R/c) G(\xi)
\label{eq:delta_I}
\end{equation}
\begin{equation}
G(\xi) = \frac{\xi \exp(\xi)}{[\exp(\xi)-1]^2} \quad {\rm where} \quad \xi = \frac{h\nu}{kT_0(R)}
\label{eq:Gxi}
\end{equation}
where $\nu$ is the frequency, $h$ is the Planck constant, and $k$ is the Boltzman constant. Equations (\ref{eq:delta_I}) and (\ref{eq:Gxi}) state that the outer edge of the disk has larger surface brightness fluctuations than the inner edge with the same temperature variation. Since the continuum reverberation mapping is only sensitive to the variable components, Equation (\ref{eq:fluxweightedR}) becomes
\begin{equation}
\langle R_\lambda \rangle = \frac{\int_{R_{min}}^{\infty} G(\xi) R^2 \, dR}{\int_{R_{min}}^{\infty} G(\xi) R \, dR} = X_{var} R_{
\lambda}
\end{equation}
where $X_{var} = 5.04$ \citep{Tie2018}. We adopt this conversion factor for further discussions unless otherwise specified.

\begin{figure*}%[ht!]
%\figurenum{1}
\gridline{\fig{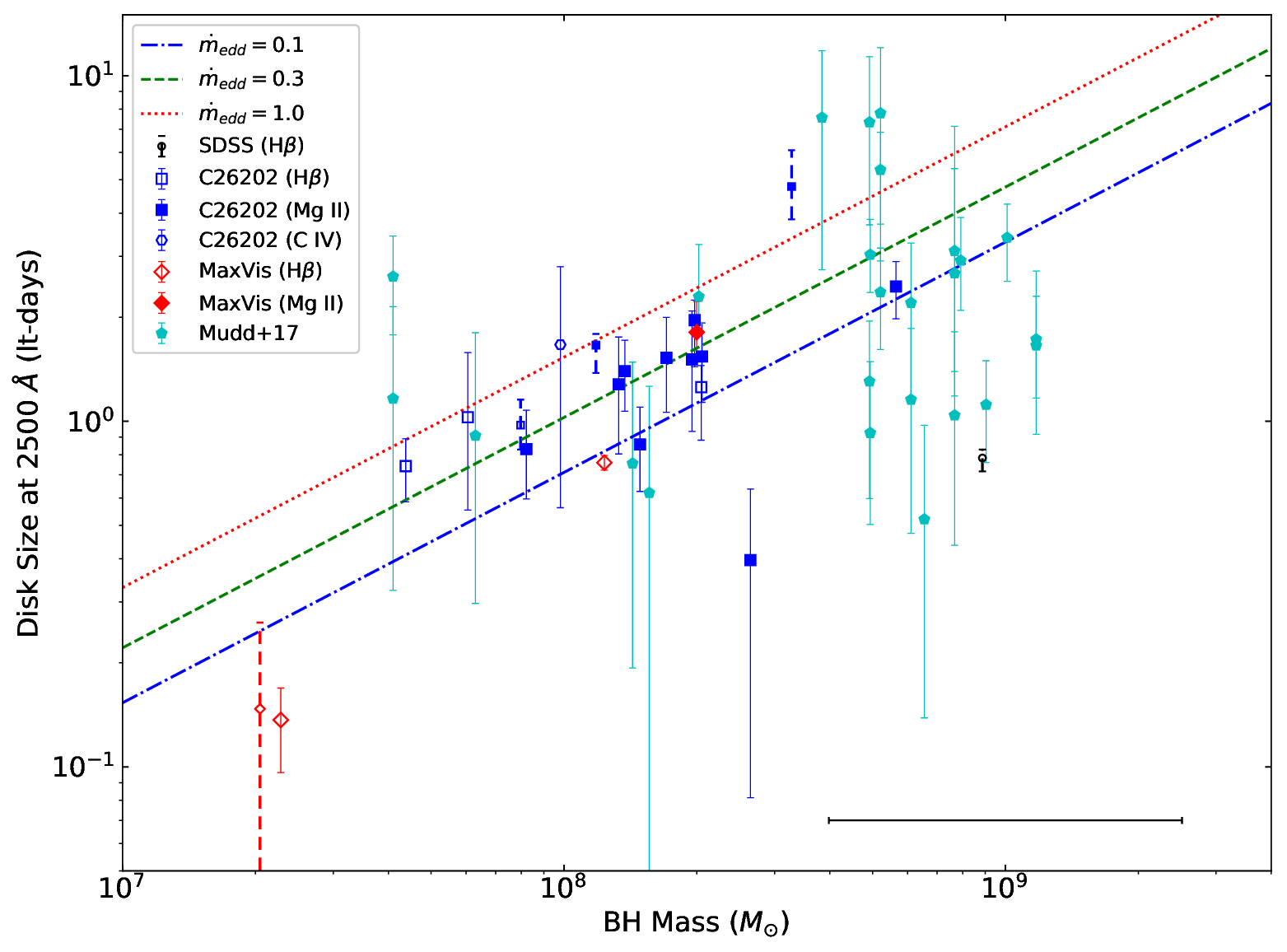}{0.7\textwidth}{}}
\figcaption{Accretion disk size at rest frame 2500 \AA \, as a function of black hole mass. The blue dash-dotted, green dashed and red dotted lines represent the prediction of the thin disk model for Eddington ratios of 0.1, 0.3 and 1.0, respectively. The red diamonds, blue squares, black circles and cyan pentagons represent the results from the MaxVis field, the C26202 field, the SDSS fields and \citet{Mudd17}, with errorbars for the 1$\sigma$ uncertainties. For the objects from this work, the smaller markers with dashed errorbars represent the flagged objects, while the larger markers with solid errorbars represent the objects with $flag=0$ (the most secure results). The open hexagons represent black hole masses measured from C IV. Other open markers represent black hole masses measured from Mg II, while the filled markers represent black hole masses measured from H$\beta$. The typical uncertainty of the black hole mass is about 0.4 dex, shown as the black errorbar in the bottom right corner. \label{fig:disksize_bhmass}}
\end{figure*}

We plot the observed accretion disk size at rest frame 2500 \AA \, for all the quasars in the main sample in Figure \ref{fig:disksize_bhmass}, and include results from \citet{Mudd17} for comparison. Many of our measurements show significantly smaller uncertainties than \citet{Mudd17}, as that paper only used the DES supernova observations with a cadence around a week, whereas we have primarily employed the standard star fields with higher observational cadences.  

\begin{figure*}%[ht!]
%\figurenum{1}
\gridline{\fig{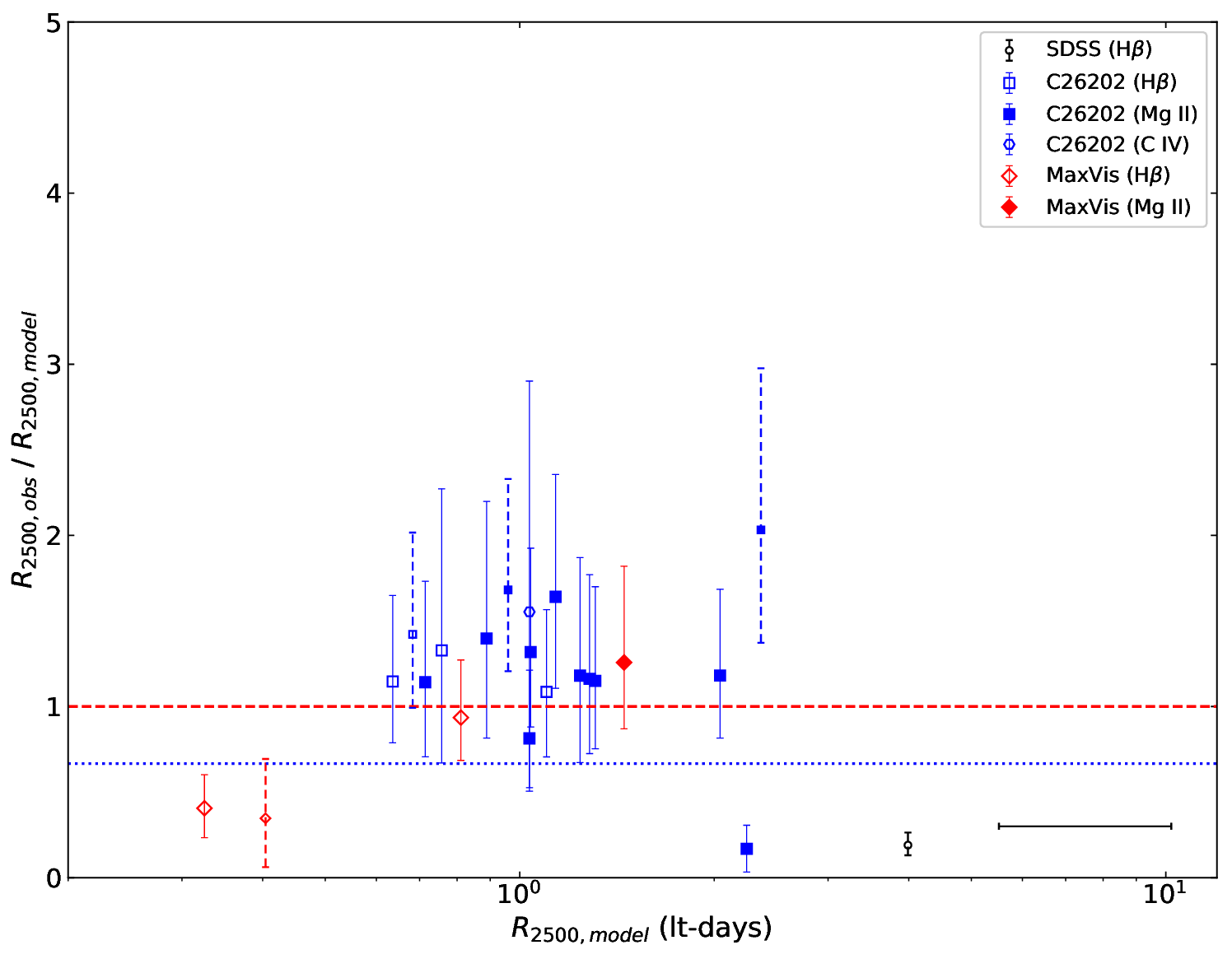}{0.7\textwidth}{}}
\figcaption{Ratio $R_{obs}/R_{model}$ of the observed disk size to the predicted disk size vs. the thin disk model prediction. The symbols have the same meaning as Figure \ref{fig:disksize_bhmass}. The red dashed line is drawn at $R_{obs}/R_{model}=1$ where the observed disk size equals the predicted disk size from the model using $X=5.04$. The blue dotted line is drawn at $R_{obs}/R_{model}=0.67$ where the observed disk size equals the predicted disk size with $X=3.36$. The uncertainty of the predicted accretion disk is shown as the black errorbar in the bottom right corner, considering that the uncertainty of the black hole mass is about 0.4 dex. \label{fig:ratio_obsvsmod}}
\end{figure*}

In addition to taking the Eddington ratio as a free parameter of the thin disk model, as is shown in Figure \ref{fig:disksize_bhmass}, we can estimate the Eddington ratio of each quasar in the main sample with the flux-calibrated OzDES spectra. We first calculate the bolometric luminosity using the continuum luminosity discussed above and the bolometric corrections $BC_{5100}=9.26$, $BC_{3000}=5.15$ and $BC_{1350}=3.81$ adopted by \citet{Shen2011} from the quasar spectra energy distributions of \citet{Richards2006}. We then divide the bolometric luminosity by the Eddington luminosity to calculate the Eddington ratio. With the Eddington ratio we calculate the predicted accretion disk size $R_{model}$ from the thin disk model adopting $X=5.04$, and the ratio $R_{obs}/R_{model}$ of the observed disk size to the predicted disk size for each quasar in the main sample. We show $R_{obs}/R_{model}$ vs. $R_{model}$ in Figure \ref{fig:ratio_obsvsmod}. Most quasars show accretion disk sizes consistent with the prediction of the standard thin disk model. This agrees with some previous studies, such as \citet{Mudd17} and \citet{Homayouni2018}. We do not find a systematic trend toward larger disk sizes as found in the Pan-STARRS sample \citep{Jiang2017} and the individual objects such as NGC 5548 \citep{Fausnaugh2016}, NGC 4151 \citep{Edelson2017} and NGC 4593 \citep{Cackett2018}. Part of the discrepancy can be due to the choice of the correction factor $X$ in the disk model. While most previous studies \citep[e.g.][]{Fausnaugh2016,Jiang2017,Cackett2018} adopted $X=2.49^{4/3}=3.36$, we adopt $X=5.04$ after taking the effect of disk variability into account. We note that if we instead use $X=3.36$ (the blue dotted line in Figure \ref{fig:ratio_obsvsmod}), the observed disk sizes for most quasars will be larger than the prediction of the model. In addition, the small number of well-studied NGC objects are among the brightest nearby AGN, and their accretion disk sizes may not be representative of typical AGN, or of the predominantly higher-luminosity quasars in our study.

%Objects without Lag Measurements
\section{Objects without Lag Measurements} \label{sec:nolagobj}
We did not recover time lags for the vast majority of the quasars that passed our first selection cuts, as described in Section \ref{sec:analysis} and illustrated in Figure \ref{fig:obs_chi2_nexp}. Only 22 quasars are in our main sample, while a total of 48 objects in the MaxVis field, 457 objects in the SDSS fields and 297 objects in the C26202 fields met the initial criteria in at least one observational season. One manifestation of a failed lag measurement is a wide, smooth lag distribution without clear peaks and centered on zero, associated with extremely large top-hat smoothing factors from JAVELIN. About 43 objects in the MaxVis field, 433 objects in the SDSS fields and 126 objects in the C26202 fields show this wide lag distribution and are excluded by Criteria (1) and (3) in Section \ref{subsec:anl-javelin}. These objects tend to have low variability amplitudes relative to the photometric uncertainties.

For quasars with significant variability, another manifestation of an unsuccessful lag measurement is a lag distribution with multiple peaks that have similar amplitudes. About 23 objects in the SDSS fields and 151 objects in the C26202 fields show lag distributions with multiple peaks and are filtered out by Criterion (2) in Section \ref{subsec:anl-javelin}. These objects can violate Criterion (1) as well if there are many secondary peaks at negative lags. While one of the multiple peaks can be physical, we cannot distinguish it from other peaks that may be caused by artifacts, so we do not include objects in the main sample if the lag distributions show multiple peaks with similar amplitudes in all bands. 

\begin{figure}%[ht!]
%\figurenum{1}
\plotone{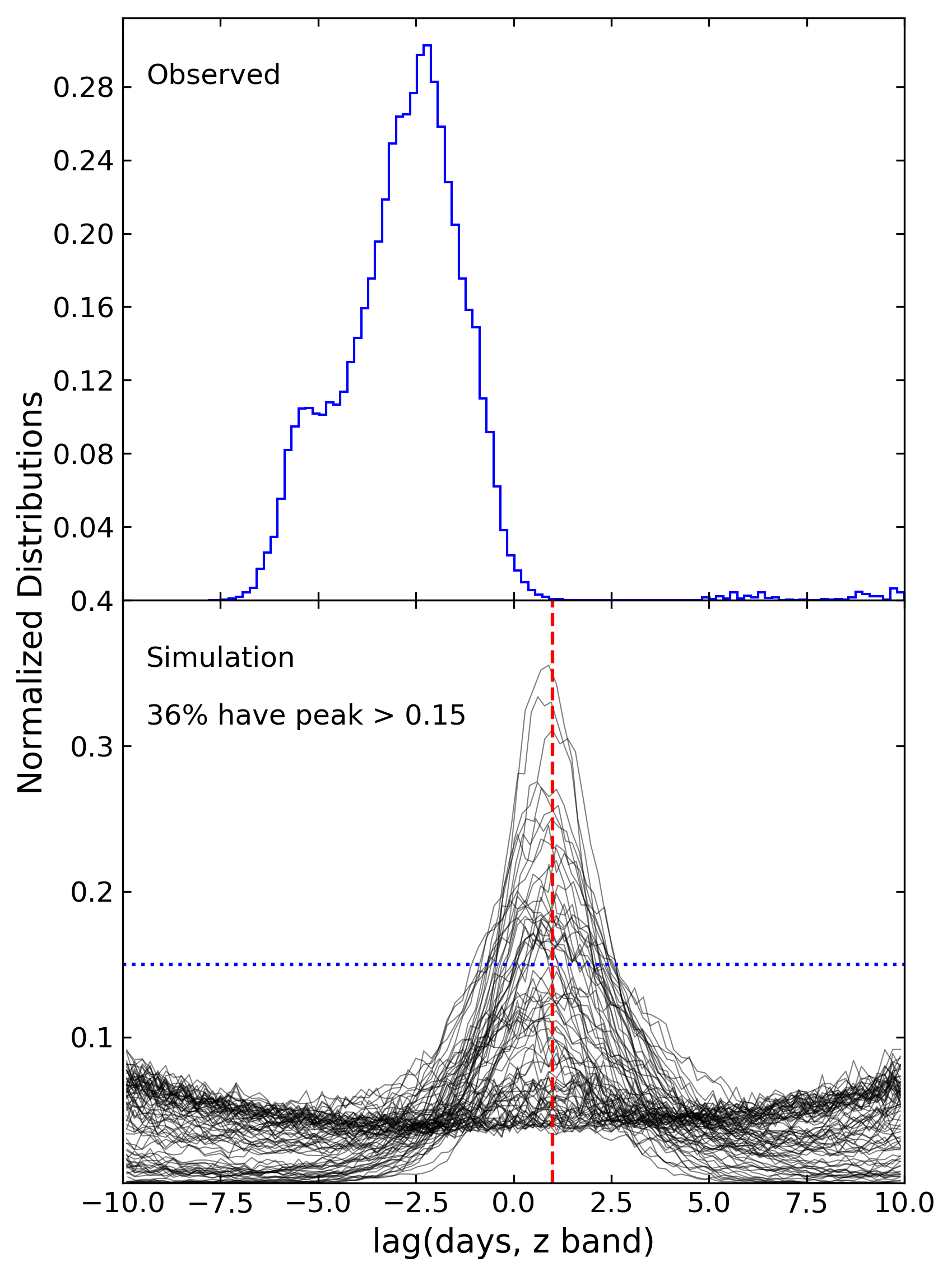}
\figcaption{(upper panel) Probability distribution of the time lags in the \textit{z} band from the observed lightcurves of DES J063227.29$-$583915.00 from JAVELIN. (lower panel) \textit{z}-band lag distributions from 80 realizations of the simulated lightcurves of DES J063227.29$-$583915.00 with an input time lag of 1 day. The red dashed line shows the position of the input time lag. The blue dotted line shows the position where the probability density equals 0.15. Only 36\% of the realizations show significant peaks with the peak probability density larger than 0.15. \label{fig:obj_zerolag}}
\end{figure}

Among the objects without good lag measurements, we find one object in the MaxVis field and two objects in the C26202 fields where JAVELIN produces clear single peaks located at zero or negative lags. As an example, the upper panel of Figure \ref{fig:obj_zerolag} shows the \textit{z}-band lag distribution from the observed lightcurves of DES J063227.29$-$583915.00. The lag distribution peaks at a negative lag, which is inconsistent with the standard thin disk model. To verify whether the negative lags are real, we perform simulations as described in Section \ref{subsec:anl-simulation} with 80 realizations. The lower panel of Figure \ref{fig:obj_zerolag} shows the simulation results of DES J063227.29$-$583915.00 for a simulated lag of 1 day, the predicted \textit{z}-band lag from the thin disk model for this object. The lag distributions from most realizations ($\sim 64\%$) are smooth distributions without clear peaks. Only a small fraction ($\sim 36\%$) show very wide peaks around the input lag with the peak probability density larger than 0.15, while nearly none of the peaks are significant lag detections according to the standards we used to generate our lag measurements of the main sample in Section \ref{sec:analysis}. This indicates that we are unlikely to obtain a reliable lag measurement for this object. We get similar results from simulations of other objects that seem to have zero or negative lag measurements, and we do not find any objects that show zero or negative lag that are verified by simulations. Note that the number of clear positive lags (22) is significantly larger than the number of possible zero/negative lags we found (3), and none of the possible zero/negative lags are as significant as the positive lags in the main sample. This indicates that the false positive rate of our sample is very low.

\begin{figure*}%[ht!]
\gridline{\fig{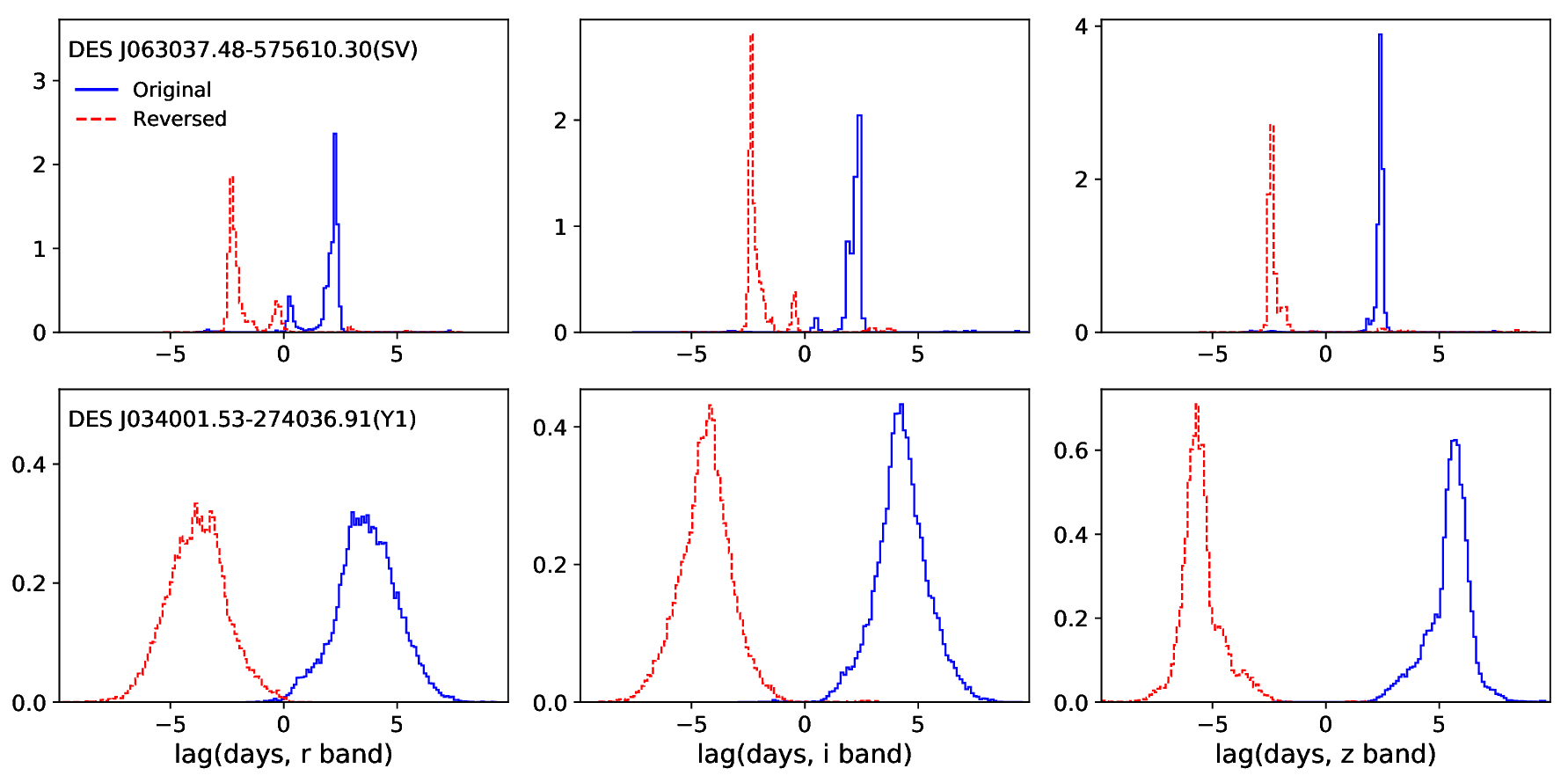}{1.0\textwidth}{}}
%\epsscale{2.0}
\figcaption{Time lag distributions from JAVELIN for the time-reversed lightcurves in comparison with the original lightcurves. The upper and lower row show the results for DES J063037.48$-$575610.30 and DES J034001.53$-$274036.91, respectively. The first through the third column give the time lag distributions in the \textit{riz} bands, respectively. The blue solid lines represents the results for the original lightcurves, while the red dashed lines are for time-reversed lightcurves. See Section \ref{sec:nolagobj} for more details.  \label{fig:rvs}}
\end{figure*}

In addition to quantifying the number of negative lags, we also reversed the time order of our lightcurves to verify that reversing the time order produces negative lags. This is to check that our lag detection algorithms are not biased to yield positive measurements. We created time-reversed lightcurves for all objects in the main sample in the seasons with lag measurements. For objects with multi-season results, we performed the test on the first season that produces lag measurements. We ran JAVELIN on these time-reversed lightcurves. Figure \ref{fig:rvs} shows examples of the comparison between the original and time-reversed results. The original and time-reversed lightcurves produce nearly \footnote{They are not exactly symmetric because of the random sampling inherent in the Markov Chain Monte Carlo method used by JAVELIN} symmetric distributions about zero. That is, reversing the time sequence simply changes the sign of the lags. In this case, the only objects that will possibly have positive detections for the time-reversed lightcurves are those with clear negative lags. As discussed above, we only found 3 insignificant zero/negative lags and therefore a very low false positive rate.

%Effect of Cadence on Lag Measurements
\section{Effect of Cadence on Lag Measurements} \label{sec:lsst}
The next generation large sky survey after DES is the Large Synoptic Survey Telescope (LSST) project \citep[e.g.][]{LSSTSciBook}. LSST will carry out a survey covering about $20000 \, {\rm deg^2}$, repeatedly scanning the region about 1000 times during a 10-year period. In addition to the main survey, LSST plans to intensively observe a set of Deep Drilling Fields (DDFs), with about 5\% of the total observing time. Each DDF has a diameter of 3.5 degrees, and the five DDFs in the LSST Reference Simulated Survey ``minion\_1016'' \citep[e.g.][]{LSSTCad} cover about $50 \, {\rm deg^2}$ in total. The large amount of observing time dedicated in each Deep Drilling Field is likely to enable the measurement of accretion disk sizes for many more quasars and with smaller uncertainties. In this section, we use our results to consider how to optimize the observation strategy for the DDFs to measure more accretion disk sizes.

One of the most important factors that affect disk size measurements is the observational cadence, which is still under discussion for the LSST DDFs. We investigate the effect of observational cadence on the fraction of quasars with good disk size measurements with simulated lightcurves. To construct simulated lightcurves, we first use the LSST Operation Simulator (OpSim) v3.3.5 \citep{OpSim} to generate the observation schedule, as well as the depth of each epoch. We use three configurations in running OpSim. One is the official configuration of the LSST Reference Simulated Survey ``minion\_1016''. In this configuration, the Deep Drilling Fields are observed on a roughly 3-day cadence in the \textit{grizY} bands. The mean $5\sigma$ depth in each epoch is about 25.02, 25.35, 24.81 and 24.48 mag for the \textit{g}, \textit{r}, \textit{i} and \textit{z} bands, respectively. We refer to this configuration as the ``3-day configuration''. We create the ``2-day configuration'' and ``1-day configuration'' by requesting OpSim to increase the number of epochs by a factor of 1.5 and 3, respectively, while reducing the exposure time in each single epoch such that the net integration time is nearly unchanged. The ``2-day configuration'' has an observational cadence around 2 days, with a median $5\sigma$ depth about 24.84, 25.13, 24.57 and 24.25 mag for the \textit{g}, \textit{r}, \textit{i} and \textit{z} bands, respectively. The ``1-day configuration'' has nearly daily observational cadence, with a median $5\sigma$ depth about 24.51, 24.73, 24.21 and 23.81 mag for the \textit{g}, \textit{r}, \textit{i} and \textit{z} bands, respectively. The total exposure time obtained in the Deep Drilling Fields for the three configurations is nearly identical. 

We then simulated a quasar with mean $(g,r,i,z) = (21.0,20.5,20.0,19.5,19.2) \, {\rm mag}$ using a DRW. We converted the magnitudes to fluxes with arbitrary units for the lightcurves, and adopted DRW parameters $\sigma_{\mbox{\tiny DRW}} = 2.5 \, {\rm (flux \: unit)}$ and $\tau_{\mbox{\tiny DRW}} = 200 \, {\rm days}$. The mean magnitudes and DRW parameters are typical of the quasars in our DES sample. We created the DRW lightcurves following a similar procedure to what is described in Section \ref{subsec:anl-simulation}, and with the cadence specified by the observation schedule from OpSim. Combining the magnitude of the quasar and the depth of LSST, we calculate the photometric uncertainties in each epoch. To allow for additional uncertainties, such as if the variability does not exactly follow the DRW, or systematics in the magnitude measurements in the single-epoch data, we set the minimum uncertainty of each data point to be 1\%, even if all the photometric uncertainties are predicted to be smaller. 

\begin{figure*}%[ht!]
\gridline{\fig{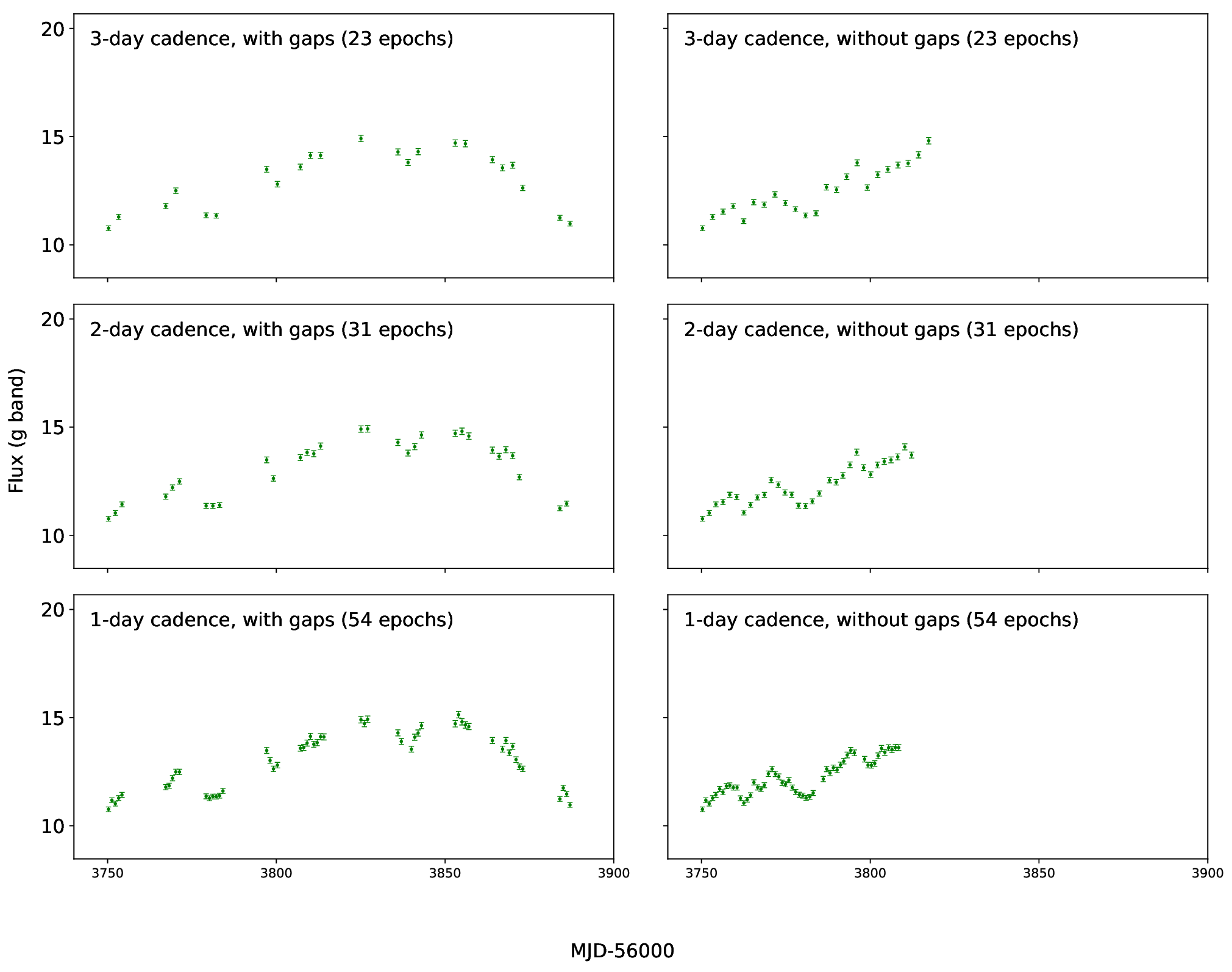}{1.0\textwidth}{}}
\figcaption{Simulated \textit{g}-band lightcurves for one observational season for quasars in the LSST DDFs. The fluxes are in arbitrary units. The upper, middle and lower row show the lightcurve from a 3-day, 2-day and 1-day cadence from OpSim, respectively. The left column shows the lightcurves with gaps, while the right column shows the lightcurves where gaps are removed. The lightcurves with the 3-day, 2-day and 1-day cadence have about 23, 31 and 54 epochs in one observational season, respectively. \label{fig:lsst_lc}}
\end{figure*}

Figure \ref{fig:lsst_lc} shows an example of the three simulated lightcurves in the \textit{g} band in the left column. One feature of the observation schedules from OpSim is that the lightcurves have gaps between a series of continuous observations, and the width of the gap is about a week. To investigate the effect of these gaps on lag measurements, we create another version of the simulated lightcurves with the gaps removed. Specifically, if an observation is more than 5 days away from the previous observation, we change this observation to the date when it is 1 day, 2 days or 3 days after the previous observation, depending on the typical cadence of the lightcurve. As part of this process, we reduce the length of the baseline, while keeping the number of epochs constant. We show an example of the lightcurves without gaps in the right column of Figure \ref{fig:lsst_lc}. These three alternate scenarios use the same total integration time, although the removal of the gaps decreases the total baseline of observations by approximately a factor of 2.

\begin{figure}%[ht!]
%\figurenum{1}
\plotone{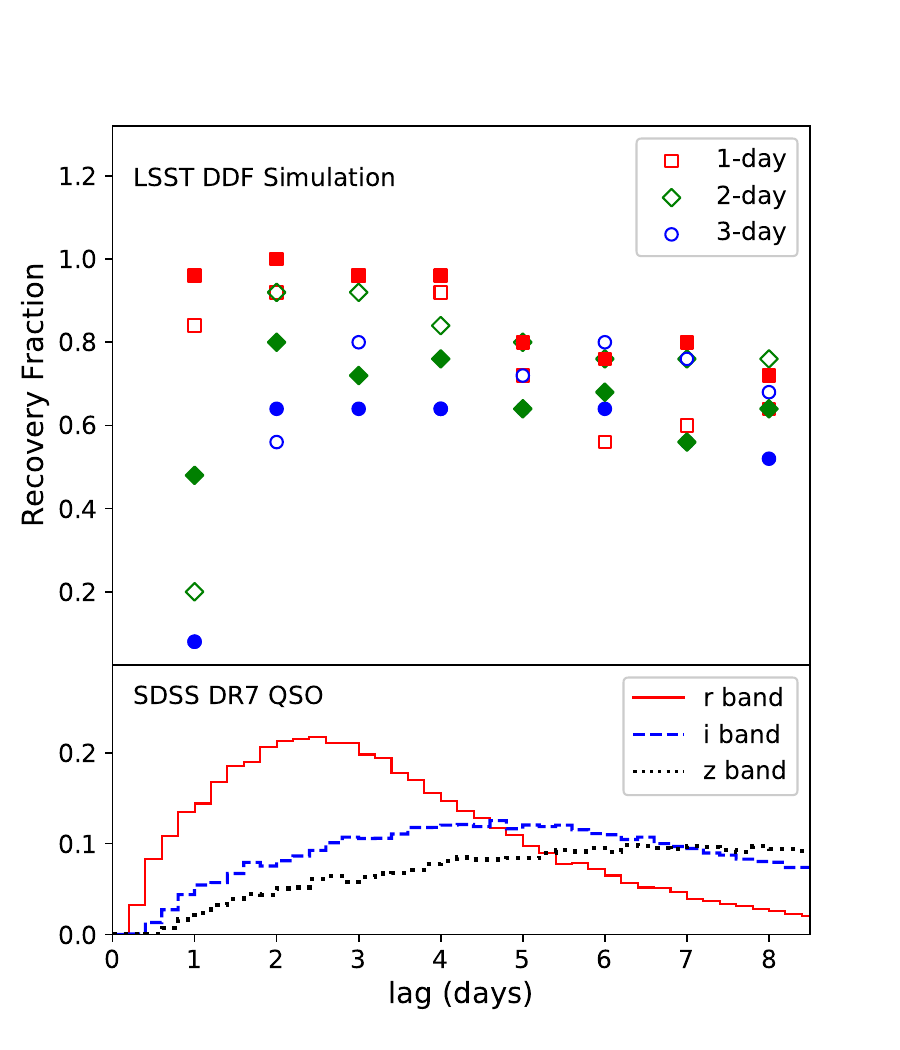}
\figcaption{(upper panel) Recovery fraction of time lags from simulated lightcurves in the LSST DDFs as a function of input time lags. The red squares, green diamonds and blue circles represent the result for lightcurves with a 1-day, 2-day and 3-day cadence, respectively. The open symbols represent results from lightcurves with gaps, while filled markers are from lightcurves without gaps. (lower panel) Predicted lag distributions for the SDSS DR7 quasars. The red, blue and black lines represents the lag distribution in the \textit{r}, \textit{i} and \textit{z} band relative to \textit{g} band, respectively. \label{fig:lsst_recfrac}}
\end{figure}

For these six scenarios, i.e. 1-day, 2-day or 3-day cadence, with or without gaps, we create 25 realizations of the DRW lightcurves in the \textit{g} band. We shift the \textit{g}-band lightcurves by the same input time lag in \textit{r}, \textit{i} and \textit{z} bands to create the simulated lightcurves in these bands. The input time lag ranges from 1 day to 8 days in steps of 1 day. We use JAVELIN to measure the time lags between simulated lightcurves, and define a successful measurement for one \textit{band} if the JAVELIN results satisfy:
\begin{enumerate}
\item The 1$\sigma$ lower limit from the probability distribution of the time lag is larger than 0;
\item The lag distribution shows a clear single peak at a positive lag; and 
\item The 1$\sigma$ upper limit of the top-hat smoothing factor is smaller than 50.
\end{enumerate}
We added the third criterion because large smoothing factors are usually associated with smooth lag distributions without clear peaks, which is a common feature of the failed fits from JAVELIN. We define a successful lag recovery for a \textit{realization} if the time lags are successfully measured in at least two of the \textit{r}, \textit{i} and \textit{z} bands. We define the lag recovery fraction as the number of realizations where we recover lags successfully divided by the total number of realizations (25). We plot the recovery fraction as a function of input time lag in Figure \ref{fig:lsst_recfrac}. For input times lags smaller than 5 days, the lag recovery fractions increase significantly toward higher cadences. Notably, for an input lag of one day, the 2-day cadence and 1-day cadence can increase the lag recovery fraction by a factor up to 5 and 10 compared to the 3-day cadence, respectively. For input time lags larger than 5 days, the trend can reverse for lightcurves with gaps, which may be due to aliasing produced by the gaps in the light curves. For lightcurves without gaps, higher cadences never produce recovery fractions smaller than those from lower cadences, as expected. This indicates that we can significantly improve the yield of accretion disk size measurements if we change the observational cadence in the DDFs from the official 3 days to 2 days or 1 day, while the total exposure time and the final coadded depth of the DDFs will not be affected. 

To estimate the distribution of the observed time lags from real quasars, we use the SDSS DR7 quasar sample presented by \citet{Shen2011}. The uniformly selected quasar sample contains 59514 quasars with flux limits of $i=19.1 \, {\rm mag}$ at $z<2.9$ and $i=20.2 \, {\rm mag}$ at $z>2.9$, where $i$ is the \textit{i}-band magnitude and $z$ is redshift. We adopt the fiducial single-epoch black hole mass from \citet{Shen2011}, and use the standard thin disk model to calculate the accretion disk size and the time lag in the \textit{r}, \textit{i}, and \textit{z} band relative to \textit{g} band assuming the Eddington ratio equals 0.1. The lower panel in Figure \ref{fig:lsst_recfrac} shows the distribution of the predicted time lag. A large fraction of the lags are less than 5 days, where higher observational cadence can significantly increase the recovery fraction. The observations of the LSST DDFs will be much deeper than SDSS, and will detect more low luminosity AGNs with smaller black hole masses and hence smaller continuum lags. In addition, AGNs with smaller black hole masses tend to be more variable and thus lag measurements should be easier. We therefore expect more small time lags in the LSST DDFs than what is implied by the lower panel in Figure \ref{fig:lsst_recfrac}. Given that the observed continuum lags from real quasars are expected to be small, a higher observational cadence in the LSST DDFs can help improve the yield of quasar accretion disk sizes significantly. 

We present an order-of-magnitude estimate of the number of successful continuum lag measurements in the LSST DDFs. We relate the number of lag measurements in the LSST DDFs to our DES results as
\begin{equation}
N_{lags} (<m_{lsst}) = f_{lsst} \times N_{qso} (<m_{lsst})
\label{eq:nlag_lsst}
\end{equation}
\begin{equation}
f_{lsst} = \frac{\Sigma_{lags} (<m_{des})}{\Sigma_{qso} (<m_{des})} \times \alpha_{cad} \times \beta_{depth}
\end{equation}
where $N_{lags,lsst}$ is the total number of AGNs with continuum lag measurements in the LSST DDFs, $\Sigma_{lags} (<m_{des})$ is the surface density of lag measurements in the DES fields, $\Sigma_{qso} (<m_{des})$ of the surface density of AGNs detected in the DES fields, $N_{qso} (<m_{lsst})$ is the number of AGNs brighter than the magnitude limit $m_{lsst}$ in the LSST DDFs, $\alpha_{cad}$ is the correction factor of different cadences between DES and LSST, and $\beta_{depth}$ is the correction factor of different depths. We define the magnitude limit $m_{lsst}$ as the magnitude corresponding to $1\%$ photometric errors and do not consider fainter objects.

The term $\Sigma_{lags} (<m_{des})/\Sigma_{qso} (<m_{des})$ represents the fraction of continuum lag measurements in the DES fields. We estimate this fraction based on the results from the C26202 fields, which combine the 1-day cadence standard star observations and 7-day cadence supernova observations. We obtain 17 good measurements out of 318 input AGNs, so we have $\Sigma_{lags} (<m_{des})/\Sigma_{qso} (<m_{des}) \approx 17/318 \approx 0.0536$.

We estimate the cadence correction factor $\alpha_{cad}$ with the same simulation framework as discussed above. The mean observational cadence of the input quasar sample in the C26202 fields is around 5 days, while the proposed cadence for AGN science in the LSST DDFs is around 2 days \citep{Brandt2018}. The mean continuum lag of the \textit{riz} bands relative to the \textit{g} band in the C26202 fields is around 3 days. We find that for an input time lag of 3 days, a 2-day cadence can increase the lag recovery fraction by a factor of about 1.35 compared to a 5-day cadence, so we have $\alpha_{cad} \approx 1.35$.

The correction factor $\beta_{depth}$ accounts for the depth difference between DES and LSST. One important difference is that deeper observations lead to smaller single-epoch uncertainties. However, for the deep DES supernova and the LSST DDF observations, the single-epoch uncertainties are dominated by calibration uncertainties instead of statistical errors, so the improvement of lag measurements from smaller single-epoch errors will not be significant. Another difference is that deeper LSST observations allows the detection of more low-luminosity AGNs, which tend to be more variable and therefore easier to measure lags. We note that the fraction of successful measurements does increase toward lower luminosities for our results from the C26202 fields, and we estimate $\beta_{depth} \approx 2.32$.

According to the LSST OpSim, the median single-epoch $5\sigma$ depth in the \textit{i} band for a 2-day cadence is about 24.57 mag, so we have $m_{i,lsst} \approx 23.8$. Based on the quasar luminosity functions by \citet{Manti2017}, we estimate ${\rm log}(N_{qso}) \approx 4.2$. Plugging these parameters to Equation (\ref{eq:nlag_lsst}), we get $N_{lags} (<m_{lsst}) \approx 2653$. That is, we expect to have about three thousand disk size measurements in the LSST DDFs with a 2-day observation cadence.

We stress that this calculation is just an order-of-magnitude estimate. Some potentially important effects that we did not include are: (1) We did not consider AGN with photometric errors larger than about 1\%, while these objects may also yield lag measurements. (2) The fraction of lag measurements based on our DES sample is limited by small number statistics. (3) When estimating $\alpha_{cad}$ and $\beta_{depth}$, we used the mean cadences, lags and depths without considering the distribution of these parameters in detail.

%Summary
\section{Summary} \label{sec:summary}
We present quasar accretion disk size measurements through continuum reverberation mapping using data from DES standard star observations. We select spectroscopically confirmed as well as color and variability selected quasars in the MaxVis field, the SDSS fields and the C26202 fields, and construct continuum lightcurves with the DES photometry from SV to Y4 in the \textit{griz} bands. We use the JAVELIN and ICCF methods to measure time lags between different bands, and use the JAVELIN Thin Disk model to fit for the accretion disk sizes. We create simulated lightcurves and re-weighted lightcurves to verify the lag measurements from JAVELIN and ICCF. We confirm that the DES photometric errors are Gaussian and appropriate for JAVELIN. We also create simulated lightcurves in the LSST DDFs, and probe the effect of observational cadence on continuum lag measurements. Our main results are:\\
\begin{enumerate}
\item We successfully measure the time lags and accretion disk sizes from 22 quasars, with black hole mass spanning $10^7$ - $10^9 \, M_{\odot}$, among which 17 have no flags and therefore are the most secure. Our measurements have smaller uncertainties than \citet{Mudd17} thanks to the higher observational cadence of the DES standard star fields.
\item Most of the measured accretion disk sizes are consistent with the predictions of the standard thin disk model if we take the disk variability into account. 
\item We have simulated several alternative observation strategies for the LSST DDFs and found that the yield of accretion disk size measurements should increase significantly if the cadence were changed from 3 days to 2 days or 1 day. 
\end{enumerate}

\bigskip

We thank the anonymous referee for a thorough and thoughtful report. This material is based upon work supported by the National Science Foundation under Grant No. 1615553. This research was funded partially by the Australian Government through the Australian Research Council through
project DP160100930. Funding for the DES Projects has been provided by the U.S. Department of Energy, the U.S. National Science Foundation, the Ministry of Science and Education of Spain, 
the Science and Technology Facilities Council of the United Kingdom, the Higher Education Funding Council for England, the National Center for Supercomputing 
Applications at the University of Illinois at Urbana-Champaign, the Kavli Institute of Cosmological Physics at the University of Chicago, 
the Center for Cosmology and Astro-Particle Physics at the Ohio State University,
the Mitchell Institute for Fundamental Physics and Astronomy at Texas A\&M University, Financiadora de Estudos e Projetos, 
Funda{\c c}{\~a}o Carlos Chagas Filho de Amparo {\`a} Pesquisa do Estado do Rio de Janeiro, Conselho Nacional de Desenvolvimento Cient{\'i}fico e Tecnol{\'o}gico and 
the Minist{\'e}rio da Ci{\^e}ncia, Tecnologia e Inova{\c c}{\~a}o, the Deutsche Forschungsgemeinschaft and the Collaborating Institutions in the Dark Energy Survey. 

The Collaborating Institutions are Argonne National Laboratory, the University of California at Santa Cruz, the University of Cambridge, Centro de Investigaciones Energ{\'e}ticas, 
Medioambientales y Tecnol{\'o}gicas-Madrid, the University of Chicago, University College London, the DES-Brazil Consortium, the University of Edinburgh, 
the Eidgen{\"o}ssische Technische Hochschule (ETH) Z{\"u}rich, 
Fermi National Accelerator Laboratory, the University of Illinois at Urbana-Champaign, the Institut de Ci{\`e}ncies de l'Espai (IEEC/CSIC), 
the Institut de F{\'i}sica d'Altes Energies, Lawrence Berkeley National Laboratory, the Ludwig-Maximilians Universit{\"a}t M{\"u}nchen and the associated Excellence Cluster Universe, 
the University of Michigan, the National Optical Astronomy Observatory, the University of Nottingham, The Ohio State University, the University of Pennsylvania, the University of Portsmouth, 
SLAC National Accelerator Laboratory, Stanford University, the University of Sussex, Texas A\&M University, and the OzDES Membership Consortium.

Based in part on observations at Cerro Tololo Inter-American Observatory, National Optical Astronomy Observatory, which is operated by the Association of 
Universities for Research in Astronomy (AURA) under a cooperative agreement with the National Science Foundation.

The DES data management system is supported by the National Science Foundation under Grant Numbers AST-1138766 and AST-1536171.
The DES participants from Spanish institutions are partially supported by MINECO under grants AYA2015-71825, ESP2015-66861, FPA2015-68048, SEV-2016-0588, SEV-2016-0597, and MDM-2015-0509, 
some of which include ERDF funds from the European Union. IFAE is partially funded by the CERCA program of the Generalitat de Catalunya.
Research leading to these results has received funding from the European Research
Council under the European Union's Seventh Framework Program (FP7/2007-2013) including ERC grant agreements 240672, 291329, and 306478.
We  acknowledge support from the Australian Research Council Centre of Excellence for All-sky Astrophysics (CAASTRO), through project number CE110001020, and the Brazilian Instituto Nacional de Ci\^encia
e Tecnologia (INCT) e-Universe (CNPq grant 465376/2014-2).

This manuscript has been authored by Fermi Research Alliance, LLC under Contract No. DE-AC02-07CH11359 with the U.S. Department of Energy, Office of Science, Office of High Energy Physics. The United States Government retains and the publisher, by accepting the article for publication, acknowledges that the United States Government retains a non-exclusive, paid-up, irrevocable, world-wide license to publish or reproduce the published form of this manuscript, or allow others to do so, for United States Government purposes.

\software{JAVELIN \citep{Zu2011}, JAVELIN Thin Disk Model \citep{Mudd17}, PyCCF \citep{pyccf}, pyspeckit \citep{pyspeckit}, Astropy \citep{astropy2013,astropy2018}, SciPy \citep{scipy}, Numpy \citep{numpy}, Matplotlib \citep{matplotlib}, Pandas \citep{pandas}.}

%Appendix
\appendix

%Comments on the lag distributions of individual objects
\section{Comments on individual lag measurements}
We comment on the time lag distributions in Figures \ref{fig:lagclean1} - \ref{fig:lagflag1} and discuss the comparisons between the JAVELIN and ICCF results for each object in the main sample. First, we clarify the ``consistency'' between the two methods. There are three cases:\\
(1) The lag distributions from both methods show clear peaks, and they agree with each other at the $1\sigma$ level. \\
(2) Only one method gives lag measurements, while the other method fails to measure the lags. This may also include the case where the lag distribution does show a peak but is much wider than the distribution from the other method.  \\
(3) The lag distributions from both methods show clear peaks, but they disagree with each other.\\
We add $flag=2$ to the objects where JAVELIN and ICCF produce inconsistent lag distributions that satisfy (2) or (3) in all bands. 

While both (2) and (3) are cases of ``inconsistent'' results, case (2) is less critical since it only indicates that the uncertainty estimates from the two methods are quite different, or one method does not work well for a particular sampling of the lightcurve. This difference between JAVELIN and ICCF has been discussed in many RM studies either on continuum or on emission lines \citep[e.g.][]{Grier2017,Mudd17,Czerny2019}, and detailed studies on this difference are out of the scope of this paper.\\ 

Comments on the unflagged objects are as follows:

\textbf{DES J063037.48-575610.30}: The ICCF peak distribution in the \textit{i} band shows a peak around 2.5 days, consistent with the JAVELIN lag distributions. The ICCF center distributions are much wider than JAVELIN. This is in fact an extreme case of the lag uncertainty difference between JAVELIN and ICCF. The ICCF center distributions do show peaks and are consistent with JAVELIN at the $1\sigma$ level, while these peaks are too wide to see clearly in normalized histogram plots. 

\textbf{DES J063510.91-585303.70}: The ICCF distributions show clear peaks in all three bands and are consistent with JAVELIN.

\textbf{DES J063159.74-590900.60}: The ICCF distributions show clear peaks in all three bands. The distributions are consistent with JAVELIN in the \textit{i} band and \textit{z} band. The main peaks of ICCF in the \textit{r} band differ slightly from the JAVELIN results, but they still agree at the $1\sigma$ level.

\textbf{DES J034003.89-264524.52}: The ICCF distributions show clear peaks in all three bands and are consistent with JAVELIN.

\textbf{DES J032724.94-274202.81}: The ICCF center distributions are much wider than JAVELIN in the \textit{r} band and \textit{i} band, i.e., the case (2) discussed above. The ICCF center distribution in the \textit{z} band shows a peak at negative lags and disagree with the JAVELIN results. However, the ICCF peak distributions show clear peaks in all three bands and are consistent with JAVELIN.

\textbf{DES J033545.58-293216.51}: The ICCF center distributions are generally much wider than JAVELIN. However, the ICCF peak distributions show clear peaks in all three bands and are consistent with JAVELIN.

\textbf{DES J033810.61-264325.00}: The ICCF distributions show clear peaks in all three bands. They are consistent with JAVELIN in the \textit{r} band and \textit{i} band. The JAVELIN results differ from the ICCF center distribution but still agree with the ICCF peak distribution at the $1\sigma$ level in the \textit{z} band.  

\textbf{DES J034001.53-274036.91}: The ICCF distributions show clear peaks in all three bands and are consistent with JAVELIN.

\textbf{DES J033853.20-261454.82}: The ICCF distributions show clear peaks in the \textit{r} band, and are consistent with JAVELIN. In the \textit{i} band the ICCF distributions are much wider than JAVELIN and are closer to the case (2) discussed above. The ICCF center distribution in the \textit{z} band is ambiguous, while the ICCF peak distribution still shows a clear peak and is consistent with JAVELIN. 

\textbf{DES J033051.45-271254.90}: The ICCF distributions are ambiguous in the \textit{r} band and \textit{z} band, which satisfies the case (2) inconsistency. However, the ICCF distribution in the \textit{i} band shows a clear peak and is consistent with JAVELIN. 

\textbf{DES J032853.99-281706.90}: The ICCF center distributions are generally much wider than JAVELIN. However, the ICCF peak distributions show clear peaks in all three bands and are consistent with JAVELIN in the \textit{i} band and \textit{z} band. The ICCF peak distributions differ from JAVELIN in the \textit{r} band.

\textbf{DES J033230.63-284750.39}: The ICCF distributions show clear peaks in all three bands and are consistent with JAVELIN.

\textbf{DES J033729.20-294917.51}: The ICCF distributions show clear peaks and are consistent with JAVELIN in the \textit{r} band. While the ICCF peak distributions are ambiguous in the \textit{i} band, the ICCF center distributions still show a peak and are consistent with JAVELIN at the $1\sigma$ level.

\textbf{DES J033220.03-285343.40}: The ICCF distributions show clear peaks in all three bands and are consistent with JAVELIN in the \textit{i} band and \textit{z} band. While the ICCF peak distributions seem to differ from JAVELIN in the \textit{r} band, they still agree at the $1\sigma$ level. 

\textbf{DES J033342.30-285955.72}: The ICCF distributions show clear peaks in all three bands and are consistent with JAVELIN, except a slight difference between JAVELIN and the ICCF center distributions in the \textit{i} band. 

\textbf{DES J033052.19-274926.80}: The ICCF distributions show clear peaks in both bands and are consistent with JAVELIN.

\textbf{DES J033238.11-273945.11}: The ICCF distributions are ambiguous in the \textit{r} band and \textit{z} band, which satisfies the case (2) inconsistency. However, the ICCF results show clear peaks in the \textit{i} band and are consistent with JAVELIN.\\

Comments on the flagged objects are as follows:

\textbf{DES J062758.99-582929.60}: The ICCF distributions are generally ambiguous and satisfy the case (2) inconsistency with JAVELIN.

\textbf{DES J033002.93-273248.30}: The ICCF distributions are generally ambiguous and satisfy the case (2) inconsistency with JAVELIN.

\textbf{DES J033408.25-274337.81}: The ICCF distribution are generally much wider than JAVELIN and are closer to the case (2) inconsistency.

\textbf{DES J032801.84-273815.72}: The ICCF distributions are generally ambiguous and satisfy the case (2) inconsistency with JAVELIN.

\textbf{DES J005905.51+000651.66}: The ICCF distributions are much wider than JAVELIN and are closer to the case (2) inconsistency in the \textit{r} band and \textit{z} band. The ICCF distributions show clear peaks in the \textit{i} band but disagree with JAVELIN.\\

In summary, all unflagged objects in the main sample have consistent JAVELIN and ICCF results in at least one of the \textit{riz} bands. Most discrepancies between JAVELIN and ICCF follow the case (2) discussed above, and the case (3) discrepancies are rare. \\

%% The reference list follows the main body and any appendices.
%% Use LaTeX's thebibliography environment to mark up your reference list.
%% Note \begin{thebibliography} is followed by an empty set of
%% curly braces.  If you forget this, LaTeX will generate the error
%% "Perhaps a missing \item?".
%%
%% thebibliography produces citations in the text using \bibitem-\cite
%% cross-referencing. Each reference is preceded by a
%% \bibitem command that defines in curly braces the KEY that corresponds
%% to the KEY in the \cite commands (see the first section above).
%% Make sure that you provide a unique KEY for every \bibitem or else the
%% paper will not LaTeX. The square brackets should contain
%% the citation text that LaTeX will insert in
%% place of the \cite commands.

%% We have used macros to produce journal name abbreviations.
%% \aastex provides a number of these for the more frequently-cited journals.
%% See the Author Guide for a list of them.

%% Note that the style of the \bibitem labels (in []) is slightly
%% different from previous examples.  The natbib system solves a host
%% of citation expression problems, but it is necessary to clearly
%% delimit the year from the author name used in the citation.
%% See the natbib documentation for more details and options.

%\begin{thebibliography}{}
%\end{thebibliography}

\bibliography{ref.bib}

%% This command is needed to show the entire author+affilation list when
%% the collaboration and author truncation commands are used.  It has to
%% go at the end of the manuscript.
%\allauthors

%% Include this line if you are using the \added, \replaced, \deleted
%% commands to see a summary list of all changes at the end of the article.
%\listofchanges

\end{document}